%% file: main.tex
\documentclass[a4paper,11pt]{article}
\pdfoutput=1 

\usepackage{jinstpub} 

\usepackage{subcaption} 
\usepackage{todonotes}  
\usepackage{isotope}    
\usepackage{siunitx}
\usepackage[version=4]{mhchem}

\title{\boldmath A 2D pixelated stilbene scintillator detector array for simultaneous radiography with fast neutrons and gammas}

\author[b,c]{R. Adams,}
\author[a]{C. G\"unther,}
\author[a,1]{N. H\"oflich,\note{Corresponding author.}}
\author[a]{O. Pooth}

\affiliation[a]{RWTH Aachen University, III. Physikalisches Institut B, Templergraben 55, 52055 Aachen, Germany}
\affiliation[b]{Swiss Federal Institute of Technology, Department of Mechanical and Process Engineering, Sonneggstrasse 3, 8092 Zurich, Switzerland}
\affiliation[c]{Paul Scherrer Institut, Nuclear Energy and Safety Research Department, 5232 Villigen PSI, Switzerland}

\emailAdd{hoeflich@physik.rwth-aachen.de}

\abstract{For radiography applications using fast neutrons simultaneously with gammas we have developed a detector with 16 stilbene crystals in a 4$\times$4 2D array with a 5~mm pitch and a depth of 25~mm. The crystal array is read out by Silicon photomultipliers and custom signal processing electronics. The detector prototype was tested using a custom D-D fast neutron generator at the Paul Scherrer Institute. By applying a pulse shape discrimination algorithm the detector is able to detect and distinguish fast neutrons and gammas simultaneously. Various attenuating samples placed between the source and detector with different materials and thicknesses were tested and the measured macroscopic fast neutron cross sections were compared to what was expected. Deviations were studied with the help of detailed Geant4 simulations. The detection efficiency for D-D fast neutrons was measured to be around 10\%. }


\arxivnumber{2010.01870} 

\begin{document}
\maketitle
\flushbottom

\section{Introduction}
\label{sec:intro}

\input{introduction.tex}
\section{Detector and set-up}
\label{sec:detector}
\input{detector_revised.tex}

\section{Experimental setup}
\label{sec:generator}
\subsection{D-D neutron source}
\label{subsec:source}
\input{source_info.tex}
\subsection{Detector arrangement}
\label{subsec:setup}
\input{measurement_setup_revised.tex}
\section{D-D generator measurement results}
\label{sec:results}
\input{results_revised.tex}

\input{geant4_revised.tex}
\section{Conclusion}
\input{conclusion_revised.tex}



\acknowledgments
We would like to thank the Paul Scherrer Institute Laboratory for Reactor Physics and Thermal-Hydraulics, in particular Heiko Kromer and Alexander Wolfertz, for the opportunity to test our detector system at their facility as well as their support during the organisation, execution and analysis of the measurements.


\end{document}

%% file: introduction.tex
Transmission radiography and tomography are powerful non-destructive testing techniques which allow visualization of the internal structure of an object. An overview of this topic is given in \cite{tomobook}. Photons and neutrons can provide complementary information depending on the materials in the object being investigated. Use of X-rays is well established, however challenges arise from their strong Z-dependence of attenuation and their typically low energy. The Z-dependence makes it difficult to get a good image contrast in low-Z materials (e.g. water) when they are surrounded by high-Z materials (e.g. a steel pipe). The typically low X-ray energies relate to much higher attenuation coefficients compared e.g. to higher energy gamma rays, limiting the size of objects which can be investigated, especially when they consist of high-Z materials. Gamma photons, e.g. 1.17~MeV and 1.33~MeV from \isotope[60]{Co}, are more penetrating, which is useful in many industrial cases where objects of interest can be large (e.g., tens of cm), however gamma imaging also suffers from the strong Z-dependence of attenuation. The higher penetrating nature of gamma rays tends to correspond to more difficulty in achieving a good detection efficiency, meaning thicker materials are preferred, and some traditional X-ray detection methods such as storage phosphor films are not suitable. Detection efficiency is of great importance for imaging with gamma rays, as typical gamma sources do not reach the output flux of X-ray sources. \\
Fast neutrons (MeV range) tend to provide even more penetration than gammas and without strong Z-dependence of attenuation. This means that, for example, steel and water (and most other materials) have macroscopic attenuation coefficients of the same order. With fast neutrons, however, unlike X-rays, convenient, compact, high flux sources do not exist. Comparatively low output neutron generators using the D-D or D-T reaction can provide a modest neutron flux for imaging. These devices typically produce X-rays in addition to neutrons, which in some neutron detector types can be a source of parasitic signals. Larger scale facilities such as reactors or spallation sources can provide higher fast neutron flux beamlines, but still with a parasitic gamma flux. Overall this means that X-ray and/or gamma photons are always present to some degree in a fast neutron imaging context. \\
The highly penetrating nature of fast neutrons, which is an advantage when imaging large objects, also creates difficulty in achieving good detection efficiency and spatial resolution for imaging. One common approach is reading out a screen consisting of a mixture of ZnS(Ag) scintillator material and polypropylene with a CCD camera \cite{zns_screen}. Another is to use an array of plastic scintillator detectors. These two options, and most other comparable alternatives, can be used to produce fast neutron images but always include a parasitic sensitivity to X-rays and gamma photons that cannot be separated from the desired contribution of fast neutrons. The usage of a scintillation material allowing the discrimination of neutron and gamma signals, such as the liquid scintillator EJ309 \cite{ej309_datasheet} or the organic crystal stilbene, allows turning the presence of gammas into added information instead of a parasitic signal. If their origin is a specific position (not distributed from all directions), simultaneous imaging with neutrons and gammas is possible, as e.g. shown in \cite{simul_rad_laptop}. \\
This work aims to achieve a position-sensitive fast neutron detector which simultaneously allows for measurement of gamma or X-ray photons of sufficiently high energy by means of pulse shape discrimination in a 2D detector array, while achieving a high detection efficiency of all particles of interest. This means that X-ray/gamma signals can either be rejected if they are parasitic or used as complementary image data. In practice typically gamma photons, not X-rays, are present in the relevant energy range, so this paper mostly refers to the higher energy photon signals only as gamma signals. The prototype presented is a 4$\times$4 pixel array. Measurements were performed to illustrate the basic performance characteristics of the prototoype when combined with a D-D fast neutron generator. D-D neutrons are useful due to their narrow and well known energy distribution. D-D neutrons have much lower energy compared e.g. to neutrons from a D-T generator. The lower energy means fewer parasitic gamma rays from inelastic neutron reactions in the detector, the investigated object, and any other surrounding materials. It also means less energy deposition in the detector on average and therefore weaker signals leading to possible challenges in neutron-gamma separability and detection efficiency.

%% file: detector_revised.tex
\subsection{The $4\times4$ pixel detector}
\begin{figure}[h!]
    \centering
    \begin{subfigure}{0.30\textwidth}
    \includegraphics[width=\textwidth]{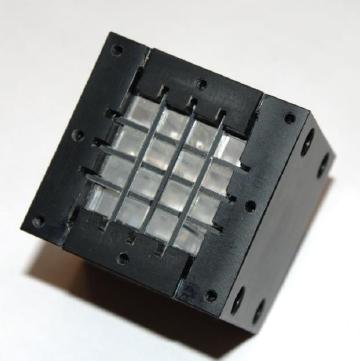}
    \subcaption{Scintillator matrix}
    \end{subfigure}
    \begin{subfigure}{0.33\textwidth}
    \includegraphics[width=\textwidth]{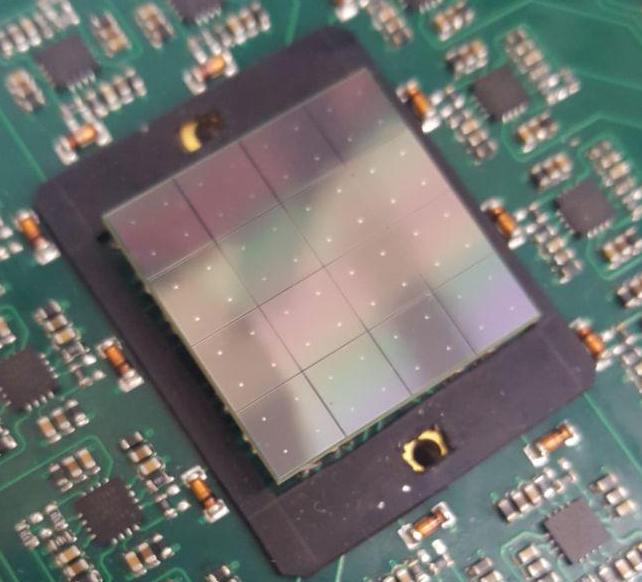}
    \subcaption{SiPM array}
    \end{subfigure}
    \begin{subfigure}{0.33\textwidth}
    \includegraphics[width=\textwidth]{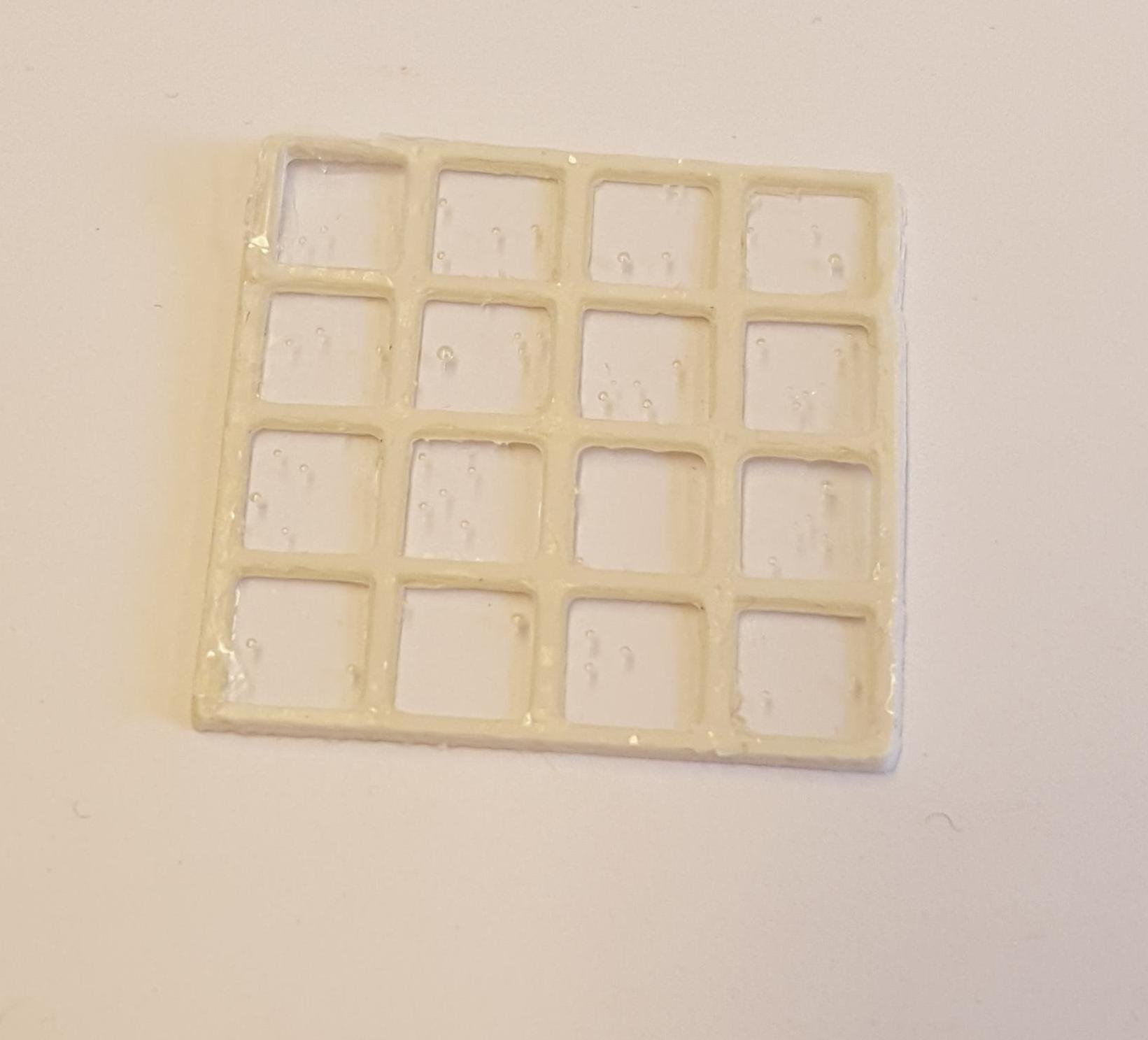}
    \subcaption{Optical coupling pad}
    \end{subfigure}
    \caption{Photographs of the $4\times4$ pixel detector components.}
    \label{fig:detector_setup}
\end{figure}

The detector, consisting of 16 stilbene scintillator cuboids of size $5 \times 5 \times 25\,\si{mm^3}$ arranged in a $4 \times 4$ array, allows the detection of neutrons and gammas with spatial resolution. The scintillators are coupled to a Hamamatsu MPPC\textsuperscript{\textregistered} S13361-6050AE-04 SiPM array \cite{sipmarray} and are separated by a grid made from black PVC\footnote{Polyvinyl chloride}. This assembly is surrounded by a black PVC housing to shield it from ambient light. All PVC surfaces facing the scintillators are covered by Vikuiti{\texttrademark} ESR (Enhanced Specular Reflecting) foil to enhance the light collection efficiency. An Elastosil\textsuperscript{\textregistered} RT~604 silicone pad is used to efficiently couple the scintillators to the SiPM array. This optical coupling pad is separated by a white PLA\footnote{Polylactic acid} grid to reduce the amount of optical crosstalk between the pixels. In Figure \ref{fig:detector_setup}, photographs of the detector components are shown. 

\subsubsection{Stilbene}
The organic scintillator crystal stilbene (chemical formula \ce{C14H12}) can be used to discriminate neutron- and gamma-induced signals. For gamma interactions, the scintillation light is created by electrons. For neutron interactions, scintillation light mainly comes from proton recoil events. Stilbene shows longer decay times of scintillation light emission for protons than for electrons \cite{stilbene_decay}. Therefore, neutron-induced signals have a longer falling edge, allowing their identification by pulse shape discrimination methods. \\
Stilbene shows a significant quenching effect. The same amount of deposited energy leads to much less scintillation light in the case of heavy charged particles (e.g. protons) compared to electrons. Also, unlike electrons, heavy charged particles exhibit a non-linear relation between deposited energy and scintillation light output \cite{stilbene_quenching_1, stilbene_quenching_2}. Anisotropy of stilbene's light response for proton recoils is reported, depending on the direction of the recoil relative to the crystal axes \cite{anisotropy}.

\subsubsection{Electronics}
\begin{figure}[h!]
    \centering
    \begin{subfigure}{0.51\textwidth}
    \includegraphics[width=1.\textwidth]{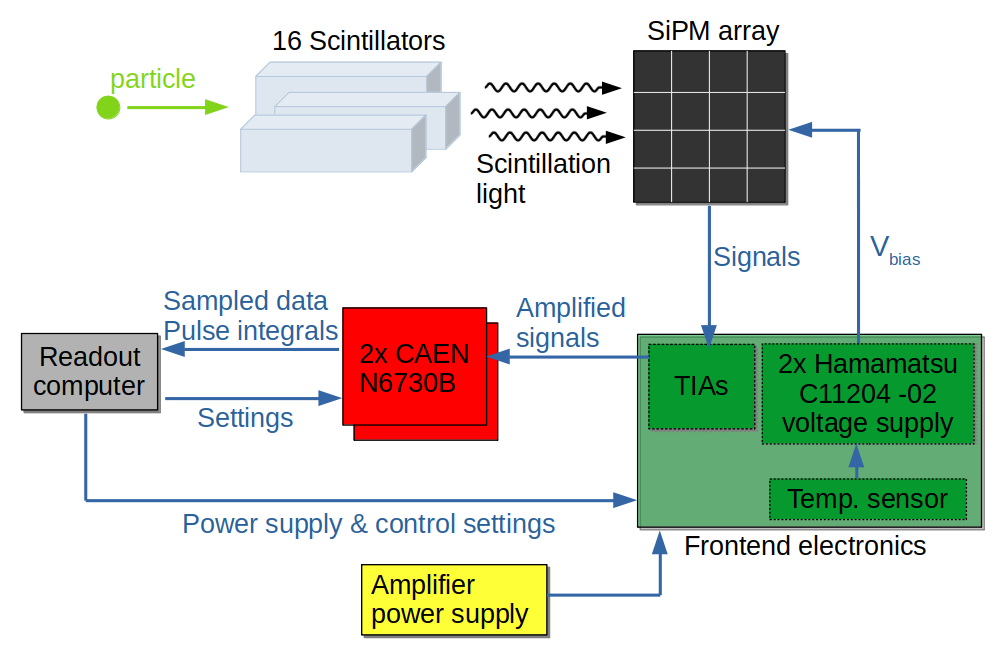}
    \subcaption{}
    \label{subfig:blockdiagram}
    \end{subfigure}
    ~
    \begin{subfigure}{0.46\textwidth}{
    \includegraphics[width=1.\textwidth]{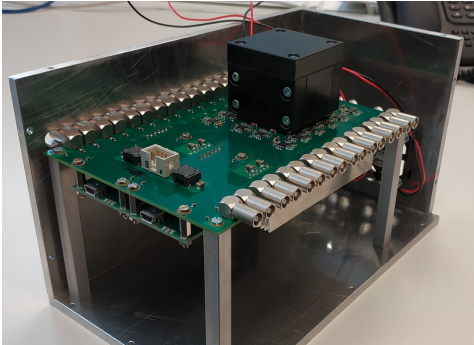}
    \subcaption{}
    \label{subfig:camerasetup}
    }
    \end{subfigure}
    \caption{(a): Schematic sketch of the detector readout. (b): Photograph of the $4 \times 4$ detector with its front end electronics.}
    \label{fig:electronics_setup}
\end{figure}

The detector readout system is shown as a schematic sketch in Figure \ref{subfig:blockdiagram}. It consists of the SiPM array, the front end electronics board developed in-house, and the digitization system. \\
The front end electronics board includes the amplifier circuits for all 16 SiPM channels as well as the SiPM bias power supply. For each channel, a positive and a negative unipolar output signal is produced to allow the use of different digitizer types. For the SiPM bias power supply, two Hamamatsu MPPC\textsuperscript{\textregistered} C11204-02 slow control chips \cite{HamamatsuPS} are used. This chip allows for a temperature correction of the SiPM's bias voltage in order to keep the SiPM gain constant. Figure \ref{subfig:camerasetup} shows a photograph of the $4 \times 4 $ detector mounted on the front end electronics board. The detector and the board are housed in an aluminium box. \\
For signal digitization, two CAEN N6730B FADCs \cite{CAENFADC} are used. They have a sampling rate of up to \SI{500}{MS \per \second}, a resolution of 14 bits and a dynamic range of \SI{2}{Vpp}.

\subsection{Pulse Shape Discrimination}
\label{sec:psd}
The classification of signals as neutron- or gamma-induced is based on the Tail-to-Total pulse shape discrimination (PSD) method. It is a commonly used method and was recently applied to measurements with stilbene in \cite{stilbene_quenching_2, psd_additional_1}. All analyses described in the following are done individually for each detector pixel. \\
The recorded signals are baseline-corrected and filtered to avoid pile-up events and parasitic signals from the microwave-driven ion source used in the neutron generator. For each remaining event, a PSD variable is calculated, defined as the ratio of the tail integral $Q_\mathrm{Tail}$ and the total integral $Q_\mathrm{L}$ of the pulse:
\begin{equation}
    PSD = \frac{Q_\mathrm{Tail}}{Q_\mathrm{L}} = \frac{Q_\mathrm{L}-Q_\mathrm{S}}{Q_\mathrm{L}}
\end{equation}
$Q_\mathrm{Tail}$ can also be expressed using $Q_\mathrm{L}$ and the short integral $Q_\mathrm{S}$ over the beginning of the pulse. The integrals are defined relative to the \textit{trigger position}, $t_\mathrm{Trg}$, where $50\%$ of the pulse height is reached, and are calculated according to:
\begin{equation}
    Q_\mathrm{L,S} = \int_{t_\mathrm{Trg} + \Delta_\mathrm{off}}^{t_\mathrm{Trg} + \Delta_\mathrm{L,S}}{S(t)\mathrm{d}t}
\end{equation}
with the signal $S(t)$. The integration limits were optimized with regard to the neutron-gamma separation to $\Delta_\mathrm{off} = -40\,\si{\nano\second}$, $\Delta_\mathrm{S} = 24\,\si{\nano\second}$ and $\Delta_\mathrm{L} = 270\,\si{\nano\second}$.  

\subsubsection{Energy calibration}
Energy calibration was performed to translate the pulse integrals of unit $\mathrm{mV \cdot ns}$ into the energy deposited in terms of $\mathrm{keV}$ electron equivalent ($\mathrm{keVee}$). For the calibration, the compton edges from a \isotope[22]{Na} and a \isotope[207]{Bi} radioactive source were used, as no photopeak is visible in the recorded gamma spectra due to the low atomic number of the elements present in stilbene. The edge positions are determined following the procedure described in \cite{calibration}. A linear relation between the pulse integral and deposited energy is assumed, based on a fit to the four edge energies and their corresponding pulse integrals. 

\subsubsection{Neutron-gamma classification and particle detection rate}
\begin{figure}[htbp!]
\centering
\begin{subfigure}{.99\textwidth}
\includegraphics[width=\textwidth]{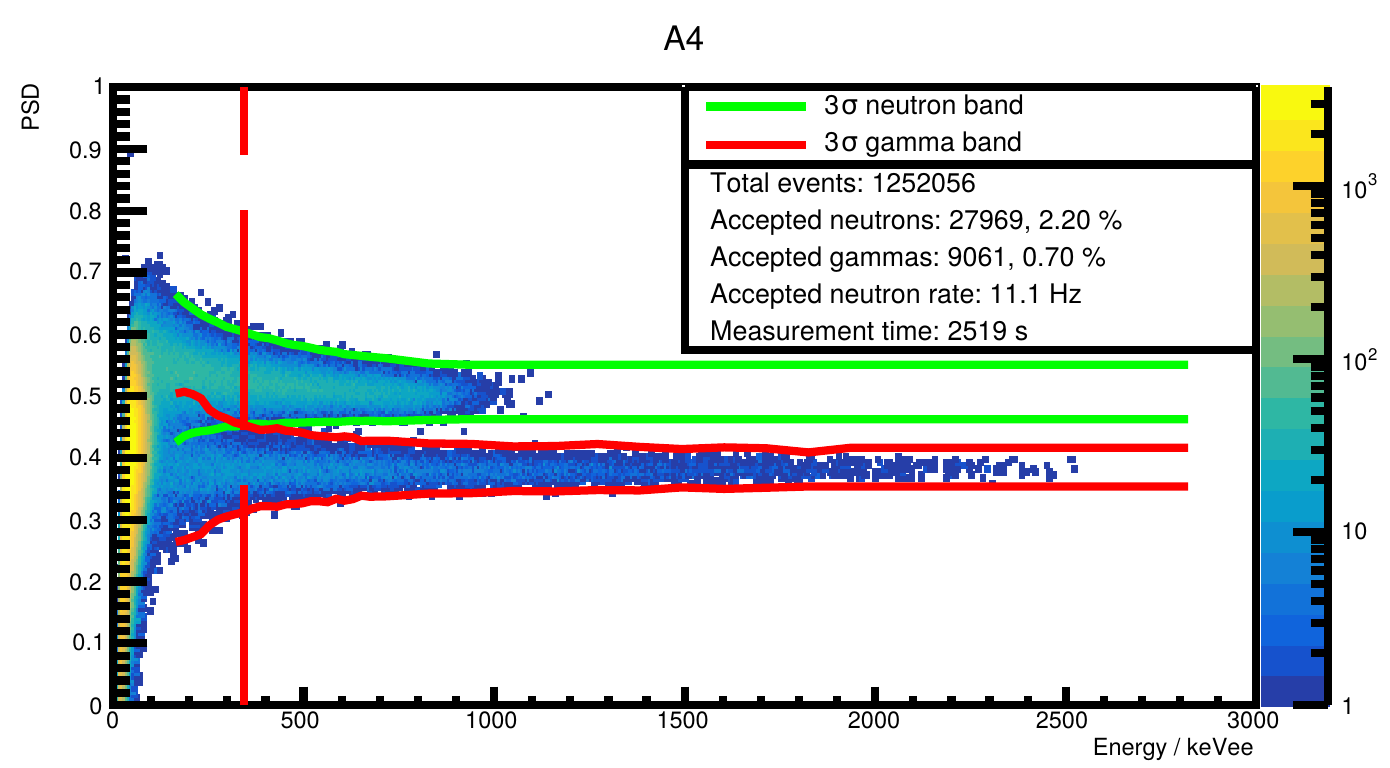}
\end{subfigure}
\caption{Example PSD plot, detector pixel A4 (pixel labeling described in figure \ref{fig:efficiency_forward}). The $3\sigma$ limits of the neutron and gamma band are shown in green and red. Extrapolations are performed to the regions where too few events for Gaussian fits are located. The red dashed line marks the energy where $FOM=1.27$.}
\label{fig:psd_3sigma}
\end{figure}

In a 2D histogram with the energy in $\mathrm{keVee}$ on the x-axis and the PSD variable on the y-axis, two distinct bands are present. An example is shown in Figure \ref{fig:psd_3sigma}, including a ``neutron band'' (upper) and a ``gamma band'' (lower). \\
To quantify the amount of neutron and gamma events, 1D histograms of the PSD variable can be produced over the range of energies using a given bin width, with the value in each bin corresponding to vertical slices from the 2D histogram. If neutron and gamma events are separated within such a 1D histogram, two peaks form which can be described by a double-Gaussian fit, corresponding to gamma events (lower peak) and neutron events (higher peak). For higher energies, almost no events occur in the neutron band region. Therefore, only the gamma band is fitted in this region. From the fit results, the neutron and gamma band limits are calculated as the $3\sigma$ boundaries of the Gaussians at each energy, later portrayed on the 2D histogram and shown in Figure \ref{fig:psd_3sigma} as green and red lines. There the neutron and gamma band limits are extrapolated horizontally into the energy regions where almost no neutron and gamma events occur. 
A criterion for the separation is the \textit{Figure of Merit (FOM)}:
\begin{equation}
    FOM = \frac{\mu_n - \mu_g}{FWHM_n + FWHM_g} = \frac{\mu_n - \mu_g}{2\sqrt{2 \ln(2)}(\sigma_n + \sigma_g)}
    \label{eq:FOM}
\end{equation}
with the means $\mu_n$ and $\mu_\gamma$ and the standard deviations $\sigma_n$ and $\sigma_\gamma$ extracted from the respective neutron and gamma band Gaussian peaks. If separation is exactly $3(\sigma_n + \sigma_g)$, the formula simplifies to $FOM = \frac{3}{2\sqrt{2 \ln(2)}}  \approx 1.27$. When calculating the FOM for all slices, one can see that it increases towards higher energies, eventually exceeding the $3\sigma$ separation value of $FOM=1.27$. In Figure \ref{fig:psd_3sigma}, the energy where $FOM=1.27$ is exceeded is marked as a red dashed line. For the region to the right of the $FOM=1.27$ position, an event is classified as neutron (gamma) if it lies inside the $3\sigma$ region of the neutron (gamma) band. The region to the left of the $FOM=1.27$ position is not taken into account due to unreliability (sensitivity to fit parameters). \\ 
Neutron and gamma rates are calculated from the particle counts in the acceptance regions divided by the measurement time. The rates, which differ from pixel to pixel, are typically around $10\,\mathrm{Hz}$ for neutrons. For the example shown in Figure \ref{fig:psd_3sigma}, the accepted neutron rate is $(11.10 \pm 0.07)\,\si{\Hz}$, corresponding to $2.2\%$ of the recorded events. The accepted gamma rate is $(3.60 \pm 0.04)\,\si{\Hz}$, corresponding to $0.7\%$ of the events. Most of the other recorded events lie at very low pulse integrals where there is no reliable separation between gammas and neutrons. \\
The $3\sigma$ limits described above were in practice determined using measurements without any object in front of the detector. In case additional measurements with objects are done with the same detector settings, the limits are re-used, as the absorption measurements are always considered in relation to the measurement without object (the so-called calibration image), see also section \ref{subsec:Absorption_meas}.

%% file: source_info.tex
\begin{figure}[htbp!]
\centering
\begin{subfigure}{.48\textwidth}
\includegraphics[width=\textwidth]{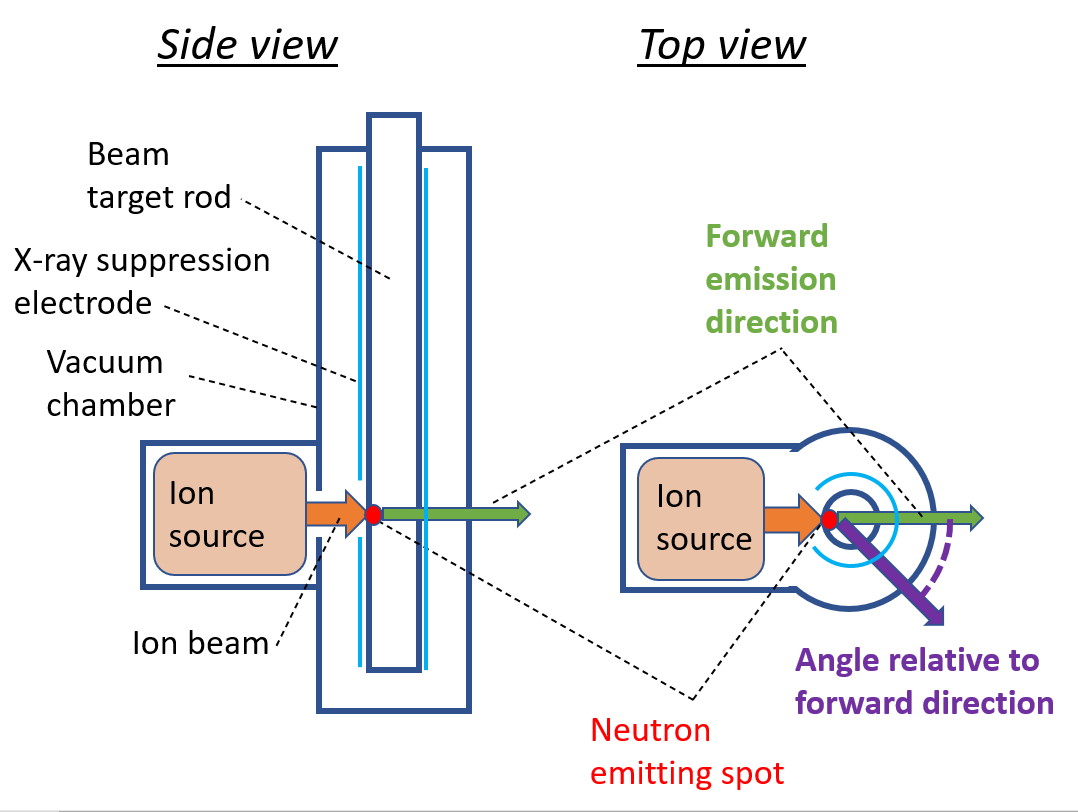}
\subcaption{}
\label{fig:source_sketch}
\end{subfigure}
~
\begin{subfigure}{.48\textwidth}
\includegraphics[width=\textwidth]{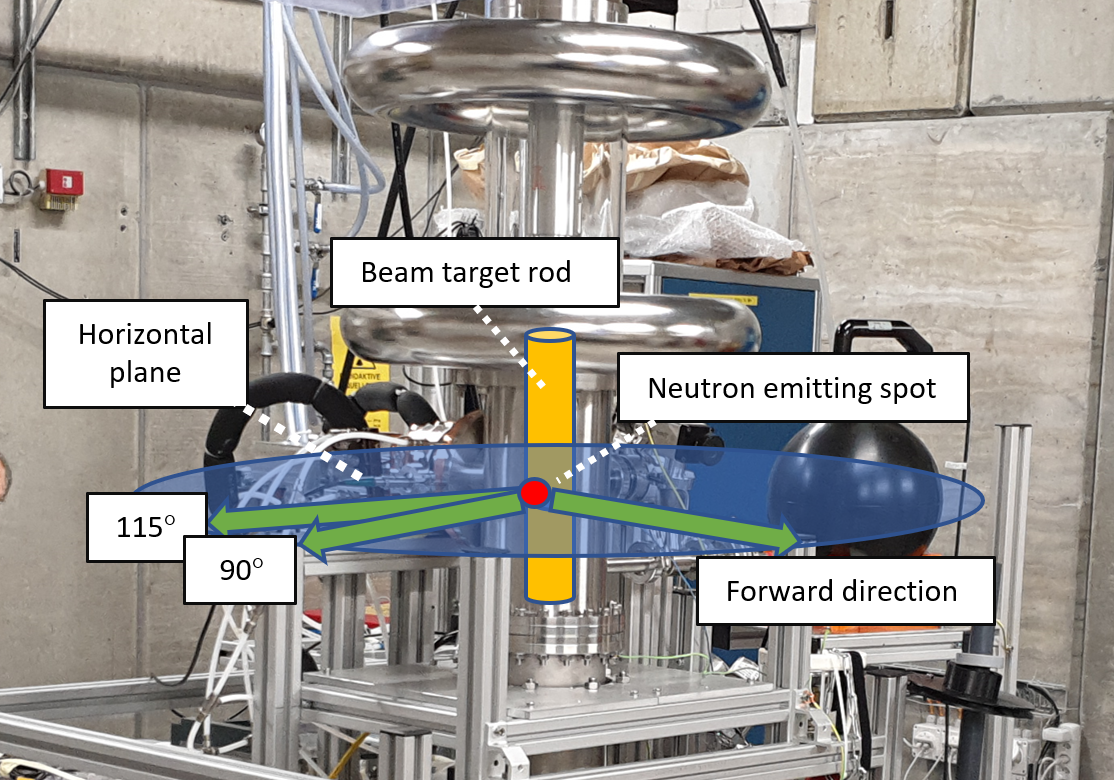}
\subcaption{}
\label{fig:source_photo}
\end{subfigure}
\caption{(a): Sketch of the PSI neutron generator with some key features labelled. (b): Photo of the PSI neutron generator with measurement directions indicated relative to the beam target rod. }
\end{figure}

This section aims to give a short overview of the compact neutron generator used in this work and some important characteristics as they relate to the measurements described in this paper. It is a custom device based on the D-D fusion reaction developed at the Paul Scherrer Institute with an emphasis on imaging, with its most notable feature being a relatively small emitting spot size of roughly 2~mm diameter intended to reduce imaging blur. The techniques used to quantify the emitting spot size, the absolute neutron output, and the flux distribution within the room where the device is operated are described in \cite{targetDesign} and \cite{sourceConfPaper}. Details about the source operating principles and design, such as the ion source and rotating ion beam target, are also described there. The deuterium ion beam loads the titanium target, ideally forming TiD$_2$ (although in reality it is likely not perfectly loaded) with which subsequent incoming deuterium ions can interact. Aside from neutrons, electrons sputtered from the target can subsequently produce Bremsstrahlung X-rays up to the acceleration energy corresponding to the voltage applied between the ion source and beam target. Efforts are made to reduce this, but not all backstreaming electrons are avoided, meaning some X-rays are always produced along with fast neutrons. A sketch of the main components of the system can be seen in Figure~\ref{fig:source_sketch}.

\noindent When a D-D fusion occurs, about half of the time it produces an alpha particle and a neutron, where the latter has an energy of 2.45~MeV in the center of mass frame. In the lab frame, the neutron energy depends on the energy and direction of the deuteron which initiates the fusion. In this work this so-called emission angle is defined relative to the nominal direction of the ion beam. For one deuterium ion direction and energy, neutrons at a given emission angle have a mono-energetic nature. Two effects cause this spectrum to become slightly polychromatic. One is that the deuterium ions do not undergo fusion at one specific energy, but over a range of energies, as they are gradually slowing down in the target rod and undergoing fusion during that slowing down process. Second, the ions are deviating in direction slightly as they slow down, meaning the nominal beam direction is not exactly the direction of a given deuteron when it undergoes fusion. These two effects were taken into consideration in detail in \cite{sourceSpectrum}.

\begin{figure}[htbp!]
\centering
\includegraphics[width=.7\textwidth]{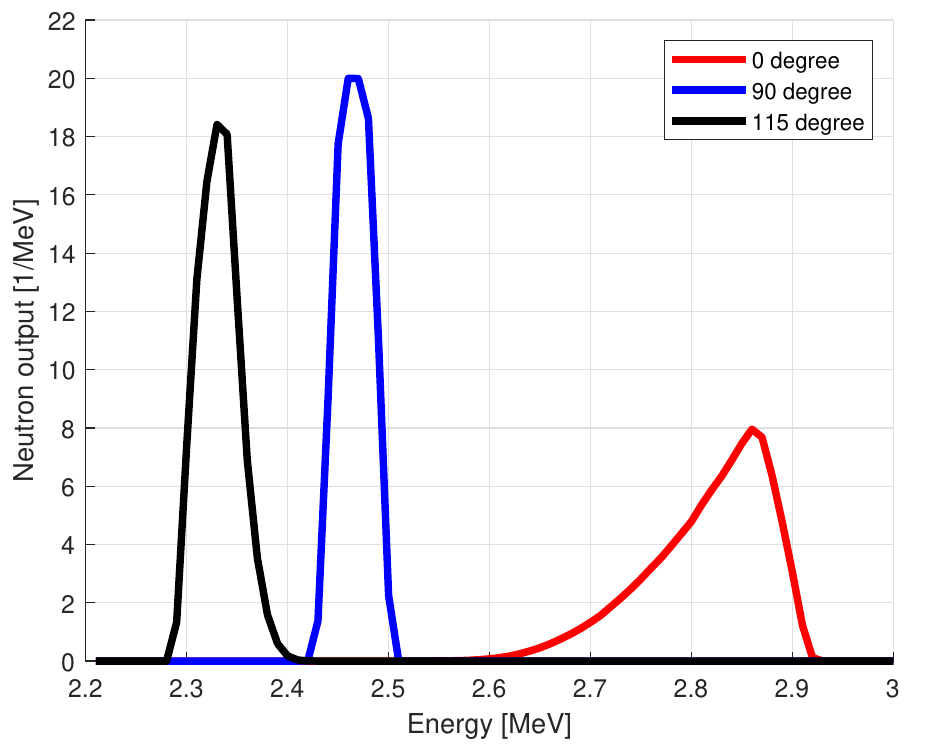}
\caption{Simulated emitted neutron spectra (normalized probability distribution function) from the PSI neutron generator at the three emission angles used for measurements at an acceleration voltage of 115~kV.}
\label{fig:emission_spectra}
\end{figure}

\noindent A photo of the neutron generator with some features indicated can be seen in Figure~\ref{fig:source_photo}. 
There the ``forward direction'' or zero degree direction is indicated which corresponds to the nominal ion beam path line. The measurements in this paper were all made in the horizontal plane at either the forward direction or one of two angles to the side (90$^{\circ}$ and 115$^{\circ}$), as indicated in the photo. The D-D reaction generally has a maximum emission energy in the forward direction and a minimum in the 180$^{\circ}$ direction. The nominal acceleration voltage used for the measurements in this paper was 115~kV. The expected neutron spectra at this acceleration voltage emitted from the source towards those three directions are shown in Figure \ref{fig:emission_spectra}. Neutrons are emitted over 4$\pi$ but with a bias towards the forward and 180$^{\circ}$ directions. At emission angles of 0$^{\circ}$, 90$^{\circ}$, and 115$^{\circ}$, the neutron rates emitted per steradian were respectively 1.45, 0.70, and 0.79 times the average over 4$\pi$ for the aforementioned acceleration voltage of 115~kV.


\noindent The output fluctuates over time but was typically about $3.5 \cdot 10^7 ~\si{\second^{-1}}$ in total over 4$\pi$. That yield was measured as a function of time using an LB6411 neutron probe. Simulations were used to
calibrate the LB6411 response relative to the absolute total neutron output, as described in \cite{{sourceConfPaper}}. The calibration enables determination of the rate of neutrons emitted at the emitting spot itself, before attenuation by nearby structures (e.g. target rod, vacuum chamber). Based on that work the neutron output is considered to have a fixed systematic uncertainty of 10\%, which does not vary from one measurement to the next. That work indicates 15\% as a maximum difference between any of the measurements performed testing the technique, and the average of those measurements, but in the present work we use 10\% according to the standard deviation of that test measurement dataset. This uncertainty is used when calculating, for example, neutron detection efficiency. However, in cases where a relative value is measured, for example from a detector response with and without an attenuating material between source and detector, this systematic uncertainty is cancelled out, leaving only stochastic uncertainty based on the finite count rate of the LB6411 detector. In the context of this work this stochastic uncertainty is often negligible.

%% file: measurement_setup_revised.tex
\begin{figure}[htbp!]
    \centering
    \begin{subfigure}{.47\textwidth}
    \includegraphics[width=\textwidth]{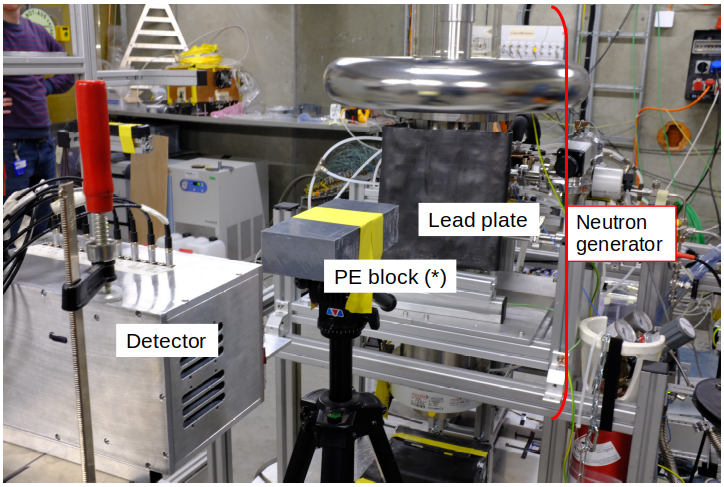}
    \subcaption{}
    \end{subfigure}
    \begin{subfigure}{.47\textwidth}
    \includegraphics[width=\textwidth]{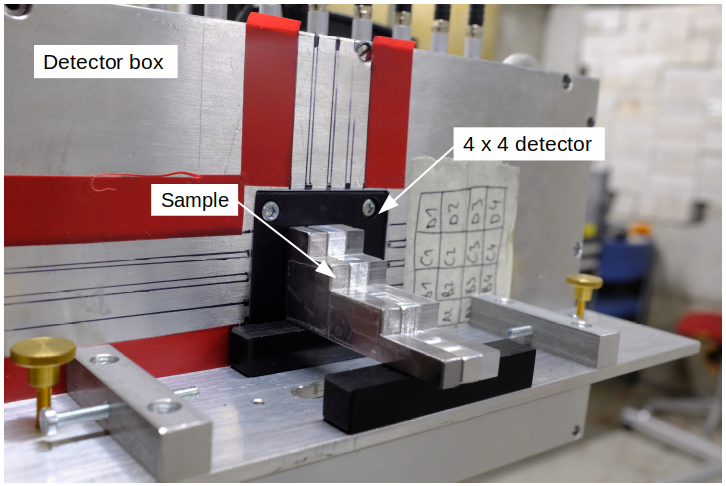}
    \subcaption{}
    \end{subfigure}
    \caption{Photographs of the measurement setup (in the forward direction). (a): Overview. (b): Detector and attenuation sample.  \textit{   (*) PE block only used for background measurements}.}
    \label{fig:setup}
\end{figure}

Measurements at \SI{0}{\degree}, \SI{90}{\degree} and \SI{115}{\degree} relative to the ion beam path were performed.
Photographs of the measurement setup are shown in Figure \ref{fig:setup}. The detector is positioned on an aluminium plate approximately \SI{80}{\centi\metre} away from the neutron emission spot. Table \ref{tab:measurement_positions} lists the exact distances and the mean neutron energies at the three measurement positions. A sample holder for small attenuating objects is attached to the detector. The objects are further described in section \ref{subsec:Absorption_meas}. Lead sheets (for reducing parasitic X-ray emission from the source) of 1~mm thickness were placed around the neutron generator which overlap at arbitrary points, resulting in an overall lead thickness which varies from 1-2~\si{\milli\metre}. For some measurements a polyethylene (PE) block of size $21 \times 10 \times 5.5~\si{cm^3}$ was placed between the generator and the detector. The PE block is expected to shield more than $99.5\%$ of the neutrons directly coming from the generator. The remaining detected neutrons are therefore considered to be ``background'' neutrons which did not travel directly from source to detector, but rather scattered off of the floor, ceiling, walls, etc. Measurement times per configuration setup varied between approximately \SI{0.5}{h} and \SI{1.0}{h}.

\begin{table}[htbp!]
    \centering
    \caption{Measurement positions and corresponding mean neutron energies. The angle is relative to the ion beam path line. The distances are determined from the neutron emission spot to the scintillators' front surface positions.}
    \label{tab:measurement_positions}
    \begin{tabular}{c|c|c}
        Angle & Mean $E_\mathrm{n}$ [MeV] & Distance [cm] \\
        \hline
        \SI{0}{\degree} & 2.84 & $81.0 \pm 2.0$ \\
        \SI{90}{\degree} & 2.47 & $81.8 \pm 2.0$ \\
        \SI{115}{\degree} & 2.33 & $79.4 \pm 2.0$ 
    \end{tabular}
\end{table}

%% file: results_revised.tex
In the following, selected measurement results obtained with the D-D generator setup are presented, focusing on the detection efficiency for D-D neutrons and neutron attenuation images of different objects. 

\subsection{Detection efficiency for D-D neutrons}
The detection efficiency $\epsilon$ is defined by 
\begin{equation}
    \epsilon = \frac{R_\mathrm{det}}{R_\mathrm{inc}}
\end{equation}
with the detected neutron rate $R_\mathrm{det}$ and the incoming neutron rate at the scintillator position $R_\mathrm{inc}$, which can be expressed by 
\begin{equation}
    R_\mathrm{inc} = R_0 \cdot \frac{F_\mathrm{det} \cdot f_\mathrm{geom} \cdot f_\mathrm{abs} }{4 \pi d^2}
\end{equation}
Here, $R_0$ is the emitted neutron rate, $d$ is the distance to the detector and $F_\mathrm{det}$ is the scintillators' front surface area. The factor $f_\mathrm{geom}$ describes the non-isotropic neutron emission of the generator, see also section \ref{sec:generator}. The factor $f_\mathrm{abs}$ describes the attenuation of neutrons in the generator itself (e.g.. vacuum chamber and target rod) for a given measurement direction, as well as in the lead plate used to shield X-rays. For the estimation of $f_\mathrm{abs}$, microscopic cross section data from the JEFF 3.3 \cite{JEFF} and ENDF/B-VII.1 \cite{ENDF} database evaluated for the neutron generator output spectrum is used. \\
The detected rate $R_\mathrm{det}$ only refers to neutrons directly coming from the generator. Therefore, the background rate, which are neutrons scattered from other directions into the detector, needs to be subtracted from the total rate measured without any object in between the generator and the detector. The background rate is determined by a measurement with PE block, as described in section \ref{subsec:setup}. To consider the fluctuating neutron generator emission, the emitted neutron rate $R_0$ is determined as the average emitted neutron rate during the measurement without object and the rate with PE block needs to be weighted by this rate so that the scattered contribution is proportionally subtracted. The weighting factor can be expressed as $R_0/R_\mathrm{0,PE}$, with $R_\mathrm{0,PE}$ being the average emitted neutron rate during the measurement with PE block.


\subsubsection{Uncertainty estimation}
The uncertainty of the efficiency consists of two parts: a statistical uncertainty from the particle counts that varies from detector pixel to pixel, and a correlated systematic uncertainty for all pixels due to a potential fixed offset between estimated and actual incoming neutron rate. This systematic uncertainty would shift the efficiencies for all pixels in the same direction.\\
The statistical uncertainty can be subdivided into a Poisson error based on the number of counts as well as an uncertainty from the $FOM=1.27$ position. The main contributions to the systematic uncertainty are the errors on the emitted neutron rate $R_0$, on the distance $d$ and on the factor $f_\mathrm{abs}$. The uncertainty on $f_\mathrm{abs}$ arises from the  uncertainties on the thickness of the lead plate, the neutron paths through the generator and, to a small extent, on the microscopic cross sections of the materials. It was roughly estimated to be a 2\% from the amount of material crossed by the neutrons in the generator structures and varying the lead plate thickness in between 1 and 2 \si{\milli\metre}. \\
The weighted mean over all pixels is calculated taking only the statistical uncertainties into account. To illustrate the effect of the correlated uncertainty, the weighted mean is also calculated after shifting the data by $1\sigma_\mathrm{sys}$ up or down.

\subsubsection{Results}
\begin{figure}[htbp!]
    \centering
    \begin{subfigure}{.68\textwidth}
    \includegraphics[width=\textwidth]{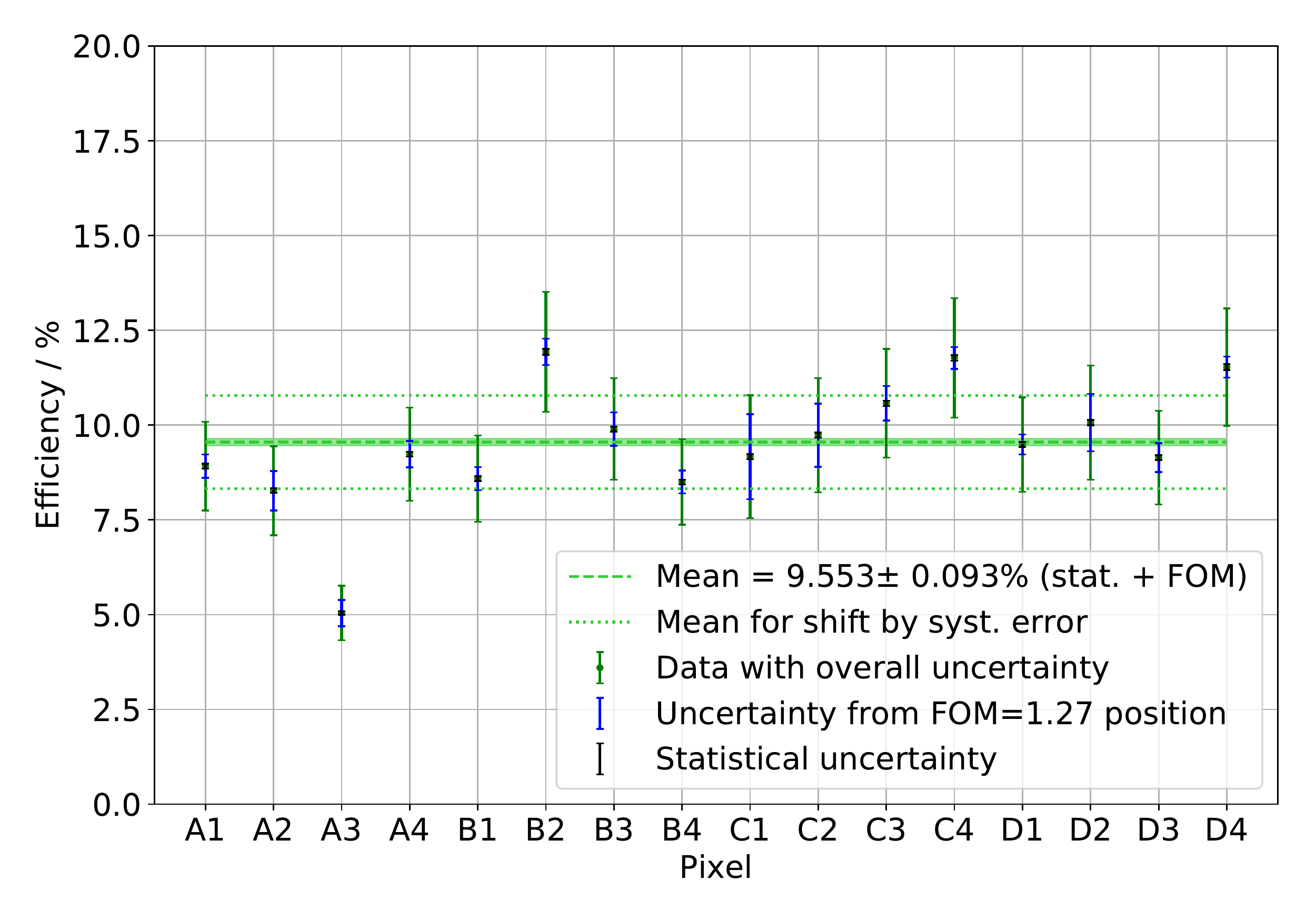}
    \end{subfigure}
    ~
    \begin{subfigure}{.3\textwidth}
    \includegraphics[width=\textwidth]{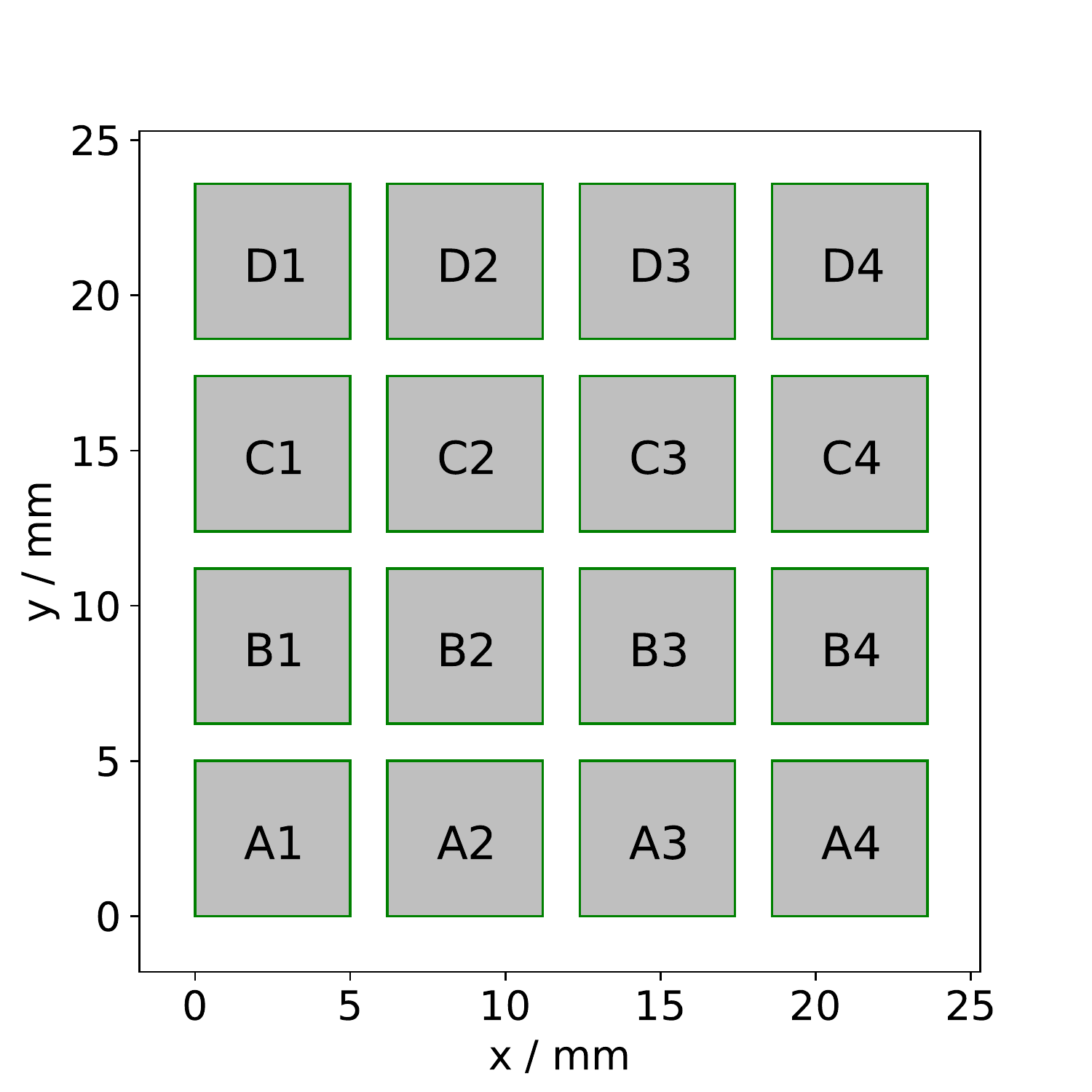}
    \end{subfigure}
    \caption{Left: Detection efficiency in \SI{0}{\degree} direction for each detector pixel. Three uncertainties are shown. The statistical (Poisson) uncertainty due to the particle counts (in black), an uncertainty due to the error on the $FOM = 1.27$ position (in blue) and a combined uncertainty including the systematic error arising from the error on $R_\mathrm{inc}$ (in green). Also, the weighted mean (green line) with its statistical and systematic error (green dotted lines) is shown. Right: Detector pixel labeling, as seen from the incoming particle's perspective.}
    \label{fig:efficiency_forward}
\end{figure}

\begin{figure}[htbp!]
\centering
\begin{subfigure}{.49\textwidth}
\includegraphics[width=\textwidth]{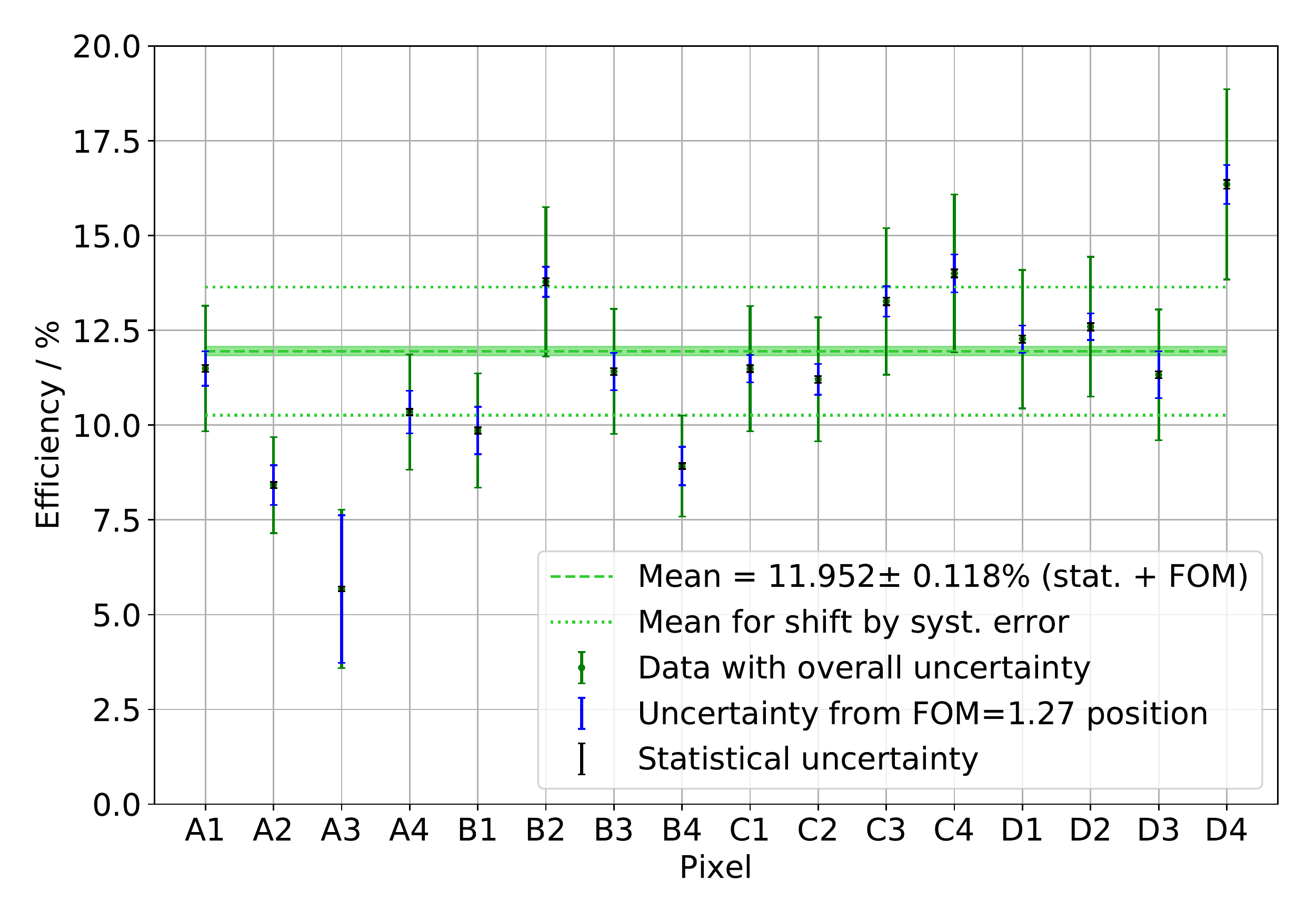}
\subcaption{\SI{90}{\degree} direction}
\end{subfigure}
~
\begin{subfigure}{.49\textwidth}
\includegraphics[width=\textwidth]{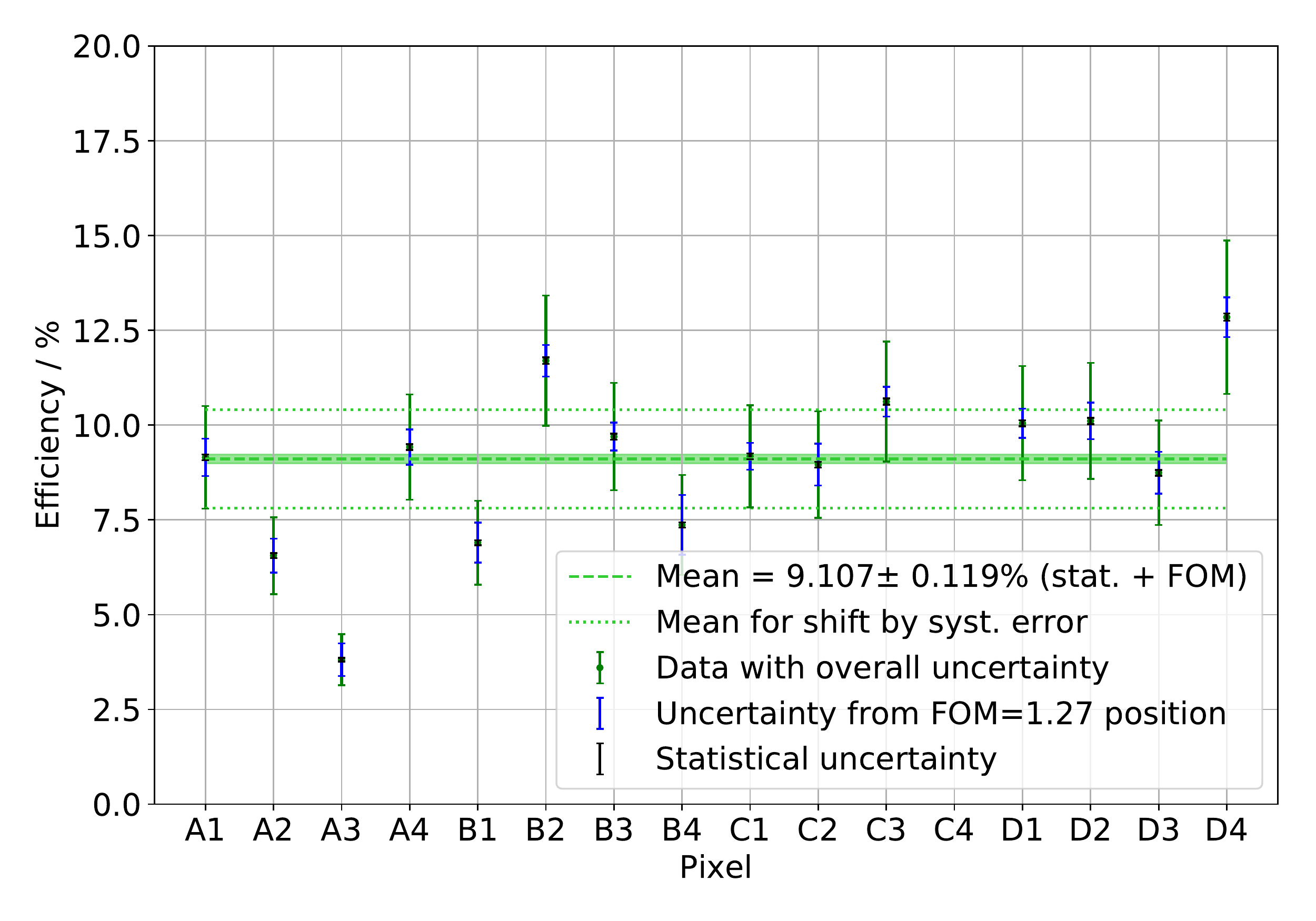}
\subcaption{\SI{115}{\degree} direction \textit{(*)}}
\end{subfigure}
\caption{Detection efficiency in \SI{90}{\degree} and \SI{115}{\degree} direction for each detector pixel. For an uncertainty description, see Figure \ref{fig:efficiency_forward}. \textit{(*): No data for pixel C4 available as it wasn't connected to the readout electronics in this measurement. }}
\label{fig:efficiency_90and115deg}
\end{figure}

In Figure \ref{fig:efficiency_forward}, the detection efficiencies for each pixel determined from a measurement in the forward direction are shown. In Figure \ref{fig:efficiency_90and115deg}, the same is shown for measurements along the \SI{90}{\degree} and the \SI{115}{\degree} directions. \\
The efficiency is overall on the order of $10\%$, but fluctuates from pixel to pixel, especially for A3 which has a much smaller efficiency than the others. A closer look at this pixel shows that its pulses are smaller and the separation between neutrons and gammas is worse than for most other pixels. In contrast, the pixels B2, C4, and D4 show an above average efficiency. In these pixels the neutron band ends at a relatively high energy, which results in better particle separation. The difference in performance from one pixel to another might be explained by variations in SiPM to crystal coupling or in characteristics of the stilbene crystal itself. Differences can occur in light propagation (e.g., different surface structures), but also in scintillation light creation, e.g., due to Stilbene's anisotropy \cite{anisotropy}.  \\
Comparing the measured efficiency in the \SI{90}{\degree} and \SI{115}{\degree} direction, the former is approximately 2.4 percentage points larger, the latter approximately 0.5 percentage points smaller than in forward direction. The variation from pixel to pixel, however, looks similar in all measurements. A larger efficiency in the \SI{90}{\degree} direction is a priori unexpected, as the neutron energy is smaller and therefore the number of events above the energy threshold should be smaller. This effect should only be partly compensated by the increase of stilbene's proton recoil interaction cross section towards lower neutron energies. The difference is probably not explained by the uncertainty in the total neutron generator output rate, as this error is expected to be a shift of the output that is in the same direction for all measurements. Also, the uncertainty on the distance and the self-absorption cannot explain the large difference. We therefore think that the difference is due to the process of neutron event identification in the two measurements. As described in section \ref{sec:psd}, only the $FOM>1.27$ energy region in the PSD plot is considered for neutron counting, and this is determined separately for each angle. For the \SI{90}{\degree} measurement, $FOM=1.27$ is reached at $E=303.2\,\si{keVee}$ when averaging over all pixels, for the forward direction, it is reached at $E=318.7\,\si{keVee}$, indicating a better separation of neutron and gamma band in \SI{90}{\degree} direction. The lower threshold needed to achieve the defined $FOM$ leads to a larger energy range considered for neutron counting, and effectively a higher detection efficiency. When applying an arbitrarily chosen constant higher energy threshold of \SI{450}{keVee} across all measured data, which is slightly above the largest $FOM=1.27$ position, the mean efficiency in the \SI{90}{\degree} direction is $4.7\,\%$, which is smaller than the forward direction value of $5.4\,\%$, which fits qualitatively much better to the expected trend. One possible explanation for the variation in neutron-gamma separation is a varying amount of gamma and X-ray flux around the source, for example if one direction was better shielded with lead than another. The lead shielding was not applied in a precisely uniform way, so it could be that the \SI{90}{\degree} direction was better lead shielded, leading to less gamma and X-ray contamination at low energies and therefore a ``cleaner'' PSD plot and a neutron-gamma separability at lower energy. However, to be certain, this should be studied more carefully in the future. 

\subsection{Attenuation measurements}
\label{subsec:Absorption_meas}
\begin{figure}[htbp!]
    \centering
    \includegraphics[width=.7\textwidth]{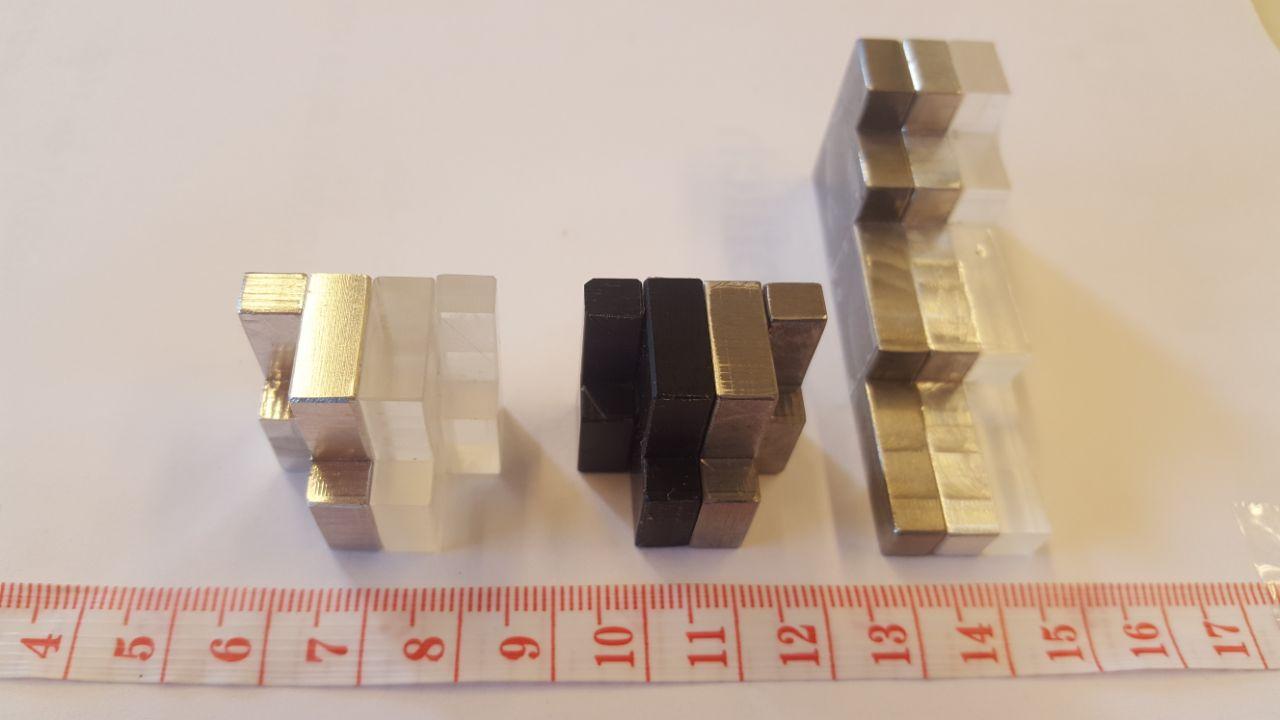}
    \caption{Photograph of the absorber objects. Left and mid: Small step wedges, step sizes $\{0.5,1.0,1.5,2.0\}\,\si{cm}$. The material combinations are Al-PMMA and PVC-Steel. Right: Large step wedge with step sizes $\{1,2,4,6\}\,\si{cm}$. The materials are steel, Al, PMMA.}
    \label{fig:absorbers}
\end{figure}

The goal of the attenuation measurements is the experimental calculation of macroscopic cross sections which can be compared to literature values. Gamma signals are separated by the PSD algorithm and gamma attenuation is not considered, as the neutron generator does not produce a localized source of gammas but they are rather created from activation processes in all surrounding materials, so that their direction of origin is not well defined. \\
Three different attenuation objects (see Figure \ref{fig:absorbers}) are investigated by placing them on a sample holder directly in front of the detector, with the detector in the forward emission direction.

\noindent Three measurements (images) were used to determine the attenuation: a so-called calibration image without any material placed in front of the camera, the so-called absorption image with the object and a background image with the PE block. In each case, a neutron rate $R_i$ with $i \in \{cal, obj, PE\}$ relative to the incoming neutron rate during each measurement is calculated. The neutron attenuation can then be expressed by the \textit{differential rate} ($\Delta Rate$) which can be defined as 

\begin{equation}
    \label{eq:DeltaRate}
    \Delta Rate = \frac{(R_\mathrm{obj} - R_\mathrm{PE}) - (R_\mathrm{cal} - R_\mathrm{PE})}{R_\mathrm{cal} - R_\mathrm{PE}} = \frac{R_\mathrm{obj} - R_\mathrm{PE}}{R_\mathrm{cal} - R_\mathrm{PE}} -1
\end{equation}

\noindent From the differential rate, the macroscopic cross section $\Sigma$ can be calculated by

\begin{equation}
   \label{eq:attenation_exp}
   R_\mathrm{obj} - R_\mathrm{PE} = (R_\mathrm{cal} - R_\mathrm{PE})\cdot e^{-\Sigma \cdot x} 
\end{equation}

\begin{equation}
    \label{eq:Sigma_exp}
     \Sigma = \frac{-1}{x} \ln{\left( \frac{R_\mathrm{obj} - R_\mathrm{PE}}{R_\mathrm{cal} - R_\mathrm{PE}}\right)} = \frac{-1}{x} \ln{\left(\Delta Rate +1\right)}
\end{equation}
The uncertainties of $\Delta Rate$ and $\Sigma$ are calculated via Gaussian error propagation from the uncertainties of $R_i$. Since the detector is not moved during a given measurement series, the uncertainties related to the detector position do not need to be considered. 

\subsubsection{Results}
\begin{figure}[htbp!]
    \centering
    \begin{subfigure}{.49\textwidth}
    \includegraphics[width=\textwidth]{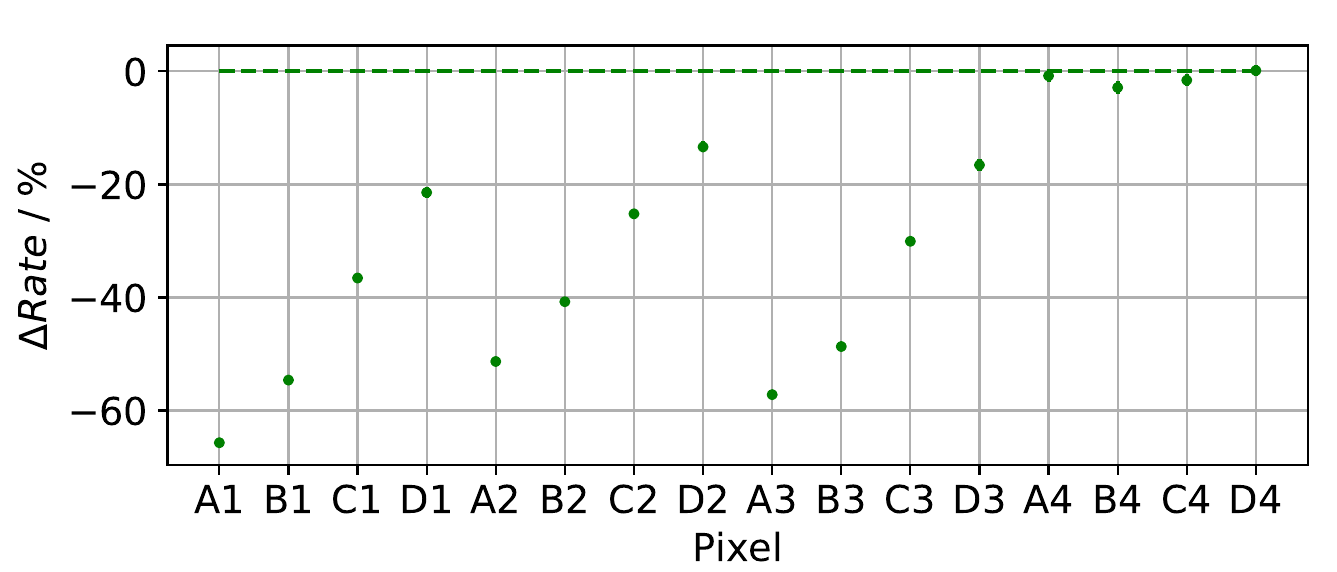}
    \subcaption{}
    \end{subfigure}
    ~
    \begin{subfigure}{.49\textwidth}
    \includegraphics[width=\textwidth]{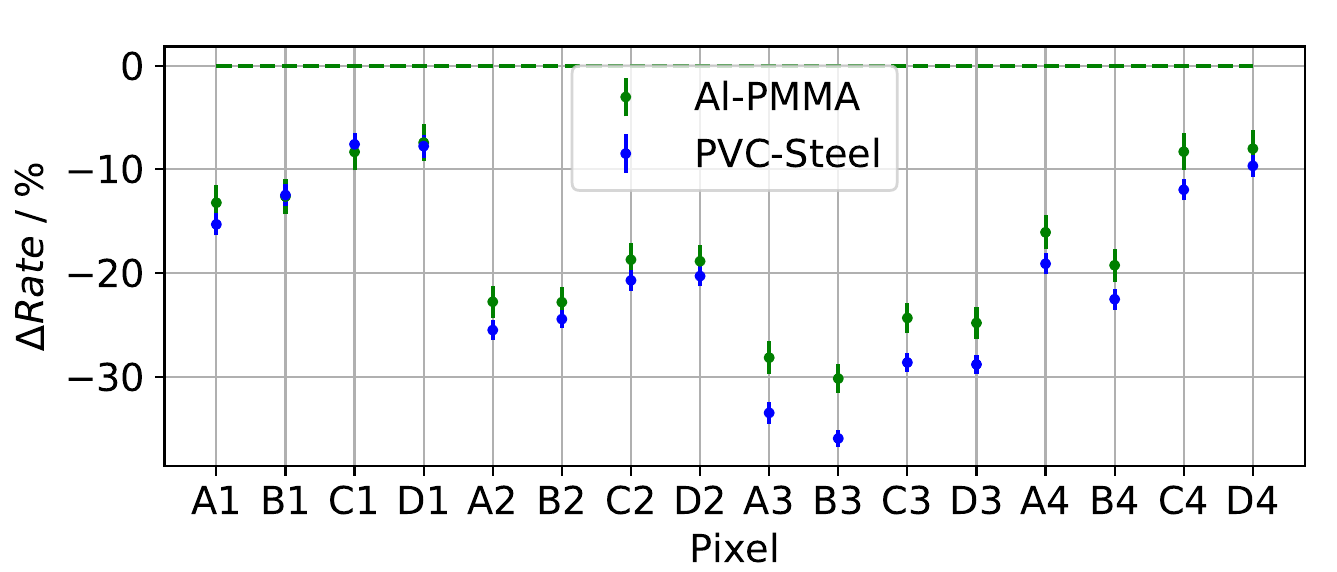}
    \subcaption{}
    \end{subfigure}
    \caption{Measured neutron differential rates for the step wedges. (a): Steel-Al-PMMA step wedge. Materials from left to right: Steel $\{6,4,2,1\}\,\si{\centi\meter}$, Al $\{6,4,2,1\}\,\si{\centi\meter}$, PMMA $\{6,4,2,1\}\,\si{\centi\meter}$, not-covered. (b): Al-PMMA and PVC-Steel step wedge. Materials from left to right: Al (PVC) $\{1,1,0.5,0.5\}\,\si{\centi\meter}$, Al (PVC) $\{2,2,1.5,1.5\}\,\si{\centi\meter}$, PMMA (steel) $\{2,2,1.5,1.5\}\,\si{\centi\meter}$, PMMA (steel) $\{1,1,0.5,0.5\}\,\si{\centi\meter}$.}
    \label{fig:diffrate}
\end{figure}

Figure \ref{fig:diffrate} shows the differential rates for the large step wedge and the small Al-PMMA\footnote{Polymethyl methacrylate} and PVC-Steel step wedges (photo in Figure \ref{fig:absorbers}). 
The large step wedge covers the detector pixel columns 1, 2 and 3 with steel, aluminium and PMMA. The pixel column 4 is left uncovered. The pixel rows A, B, C and D are covered by \SI{6}{cm}, \SI{4}{cm}, \SI{2}{cm} and \SI{1}{cm} of material, respectively (detector pixel arrangement in Figure \ref{fig:efficiency_forward}). Looking at the differential rates, one can clearly identify the step wedge's structure. Also, differences between the materials can be seen. The smallest attenuation is seen for \SI{1}{\centi\metre} aluminium, the largest attenuation for \SI{6}{\centi \metre} steel. Overall, aluminium is the weakest attenuator and steel is the strongest. For the uncovered row, no significant rate reduction is seen, indicating in particular no shadowing effect. \\
The small step wedges cover all detector pixels. Always two neighbouring pixels in the vertical direction are covered by the same material thickness. The pixel columns 1 and 2 are covered with Al (PVC), the columns 3 and 4 with PMMA (steel). Pixel rows C and D are covered by \SI{0.5}{\centi \meter} or  \SI{1.5}{\centi \meter}, A and B by  \SI{1}{\centi \metre} or  \SI{2}{\centi \metre} of material. In the differential rates, these small thickness differences are still visible, despite some fluctuations. As for the large step wedge, aluminium is the weakest attenuator, it attenuates much less than PMMA and slightly less than PVC. The strongest attenuator is steel. However, the differential rates also show a characteristic problem of the radiographic imaging method. For example, $\Delta Rate$ for \SI{1.5}{\centi \meter} Al and \SI{1}{\centi \meter} PMMA are comparable, so, for an unknown sample, it would be impossible to discriminate between two such objects in case both material and thickness are unknown. Only if one of the quantities is known a priori, the other can be potentially determied from a radiographic image, as long as the materials to be identified have sufficiently different attenuation coefficients. With a tomographic image, this problem is not present, and a spatial distribution of the attenuation coefficient could be determined.  

\begin{figure}[htbp!]
    \centering
    \begin{subfigure}{.49\textwidth}
    \includegraphics[width=\textwidth]{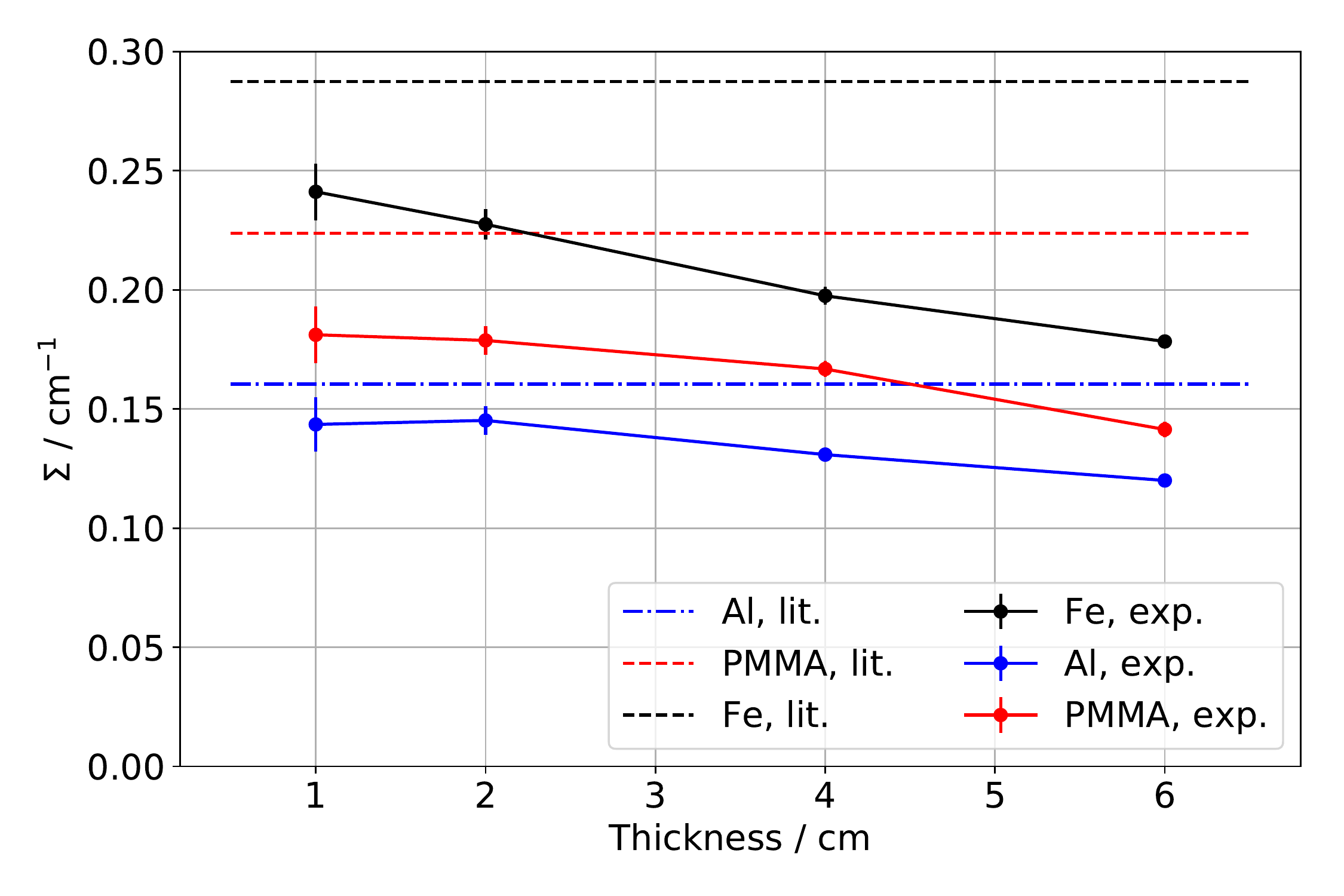}
    \subcaption{Step wedge Steel-Al-PMMA}
    \end{subfigure}
    ~
    \begin{subfigure}{.49\textwidth}
    \includegraphics[width=\textwidth]{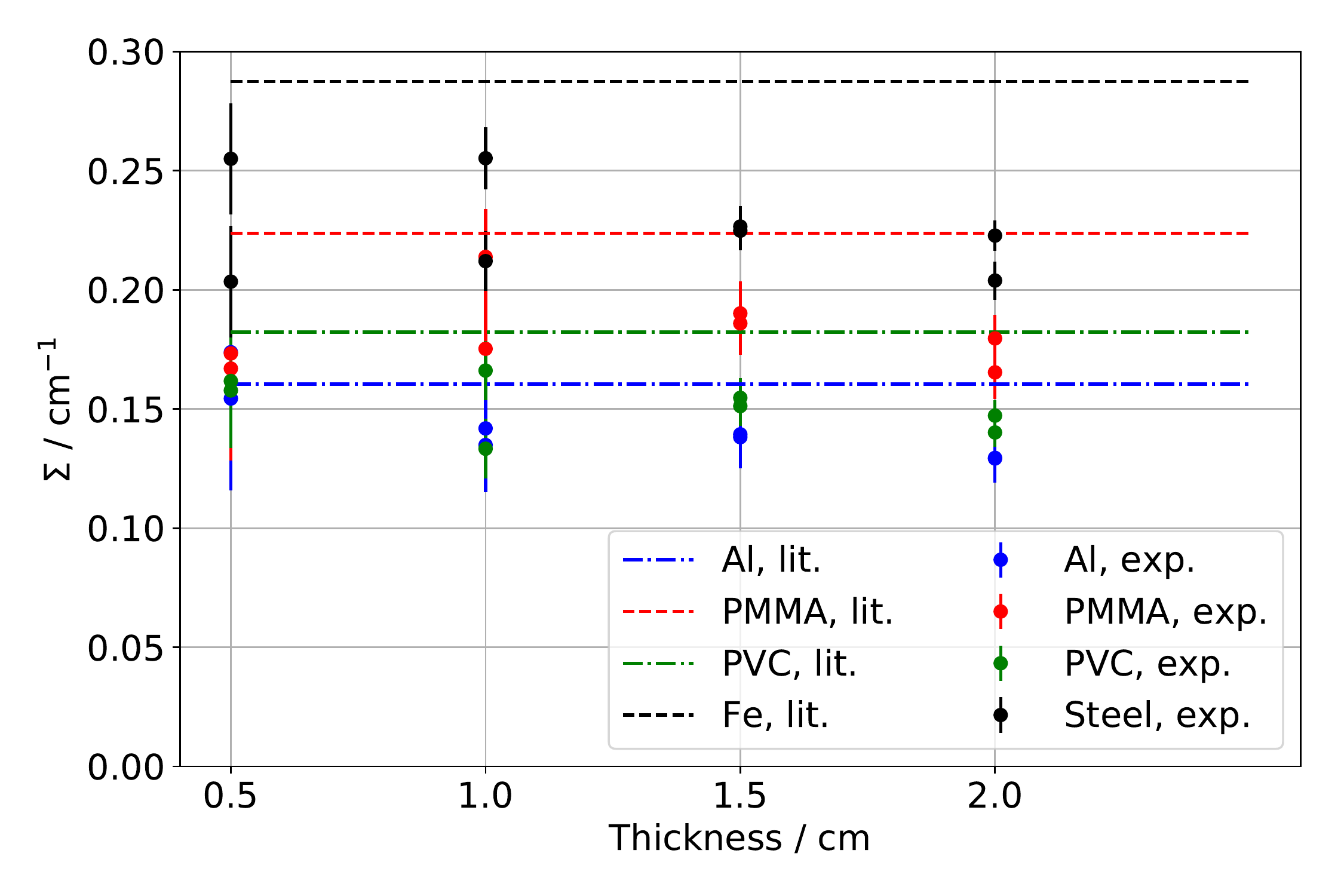}
    \subcaption{Step wedges Al-PMMA and PVC-Steel}
    \end{subfigure}
    \caption{Macroscopic cross section for steel, aluminium, PMMA, and PVC, determined from measurements with the Fe-Al-PMMA step wedge (in (a)) and the small step wedges (in (b)). The literature values calculated from JEFF 3.3 and ENDF/B-VII.1 microscopic cross section data are marked in the plots as dotted lines. }
    \label{fig:mac_crosssection}
\end{figure}

\noindent From the $\Delta Rate$, the macroscopic cross sections are calculated according to equation \eqref{eq:Sigma_exp} and are shown in Figure \ref{fig:mac_crosssection} both for the large step wedge and the small step wedges Al-PMMA and PVC-Steel. The figures also show the expected values according to microscopic cross section data (JEFF3.3 and ENDF/B-VII.1 database) combined with the previously given fast neutron spectra. The results from the large and the small step wedges for steel, Al and PMMA are within the uncertainties. While the material sequence is correctly reproduced within the uncertainties, it can be seen that the experimentally determined cross sections are generally significantly smaller that the literature values. For the large step wedge, the difference gets larger as the material thickness increases, especially for steel. This discrepancy was further studied using simulations, as described in section \ref{sec:Geant4}.

%% file: geant4_revised.tex
\subsection{Comparison with Geant4 simulations}
\label{sec:Geant4}
Geant4 simulations of the setup were done to better understand the neutron interactions in the attenuator objects and in the detector. 

\subsubsection{Geant4 setup}
\begin{figure}[htbp!]
    \centering
    \includegraphics[width=0.7\textwidth]{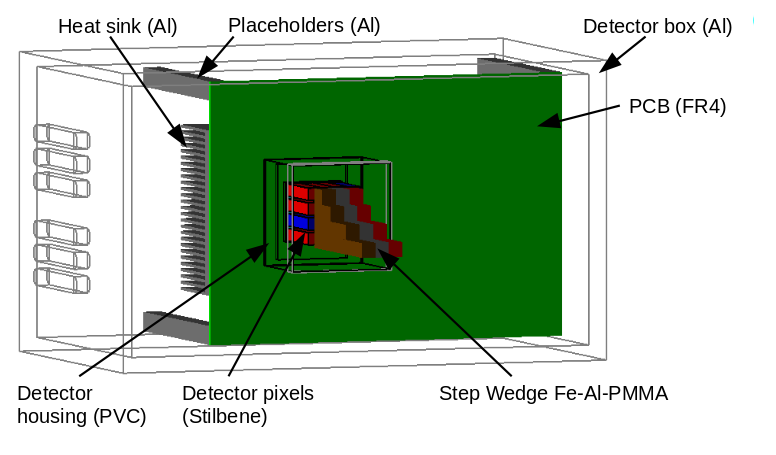}
    \caption{Geant4 detector model with attenuator step wedge (Fe-Al-PMMA) placed in front of the detector. In the simulation, steel is approximated as pure iron because the exact steel composition is unknown, but the difference in attenuation should be very small.}
    \label{fig:geant4model}
\end{figure}

The Geant4 simulation setup consists of a detailed detector model along with attenuator objects and optionally a PE block, as shown in Figure \ref{fig:geant4model}. It is assumed that incoming neutrons are emitted from a point source at \SI{80}{cm} distance from the detector with a spectrum defined according to an emission in either the forward or \SI{90}{\degree} direction. \\
The simulation stores the neutron interactions in the step wedge and the interactions, energy losses by particle type and amount of created scintillation light in the detector pixels. 
Geant4 version 10.05.p01 \cite{Geant4} is used and the physics list essentially corresponds to the QGSP\textunderscore BIC\textunderscore HP list, but using G4EmStandardPhysics\textunderscore option4 and G4OpticalPhysics \cite{Geant4Physics}. The JEFF3.3 library is used for the neutron interaction data. The particle-dependent scintillation light output is modeled based on Birk's law \cite{Birk}. 

\subsubsection{Spectra and Efficiency}
\begin{figure}[htbp!]
    \centering
    \includegraphics[width=.49\textwidth]{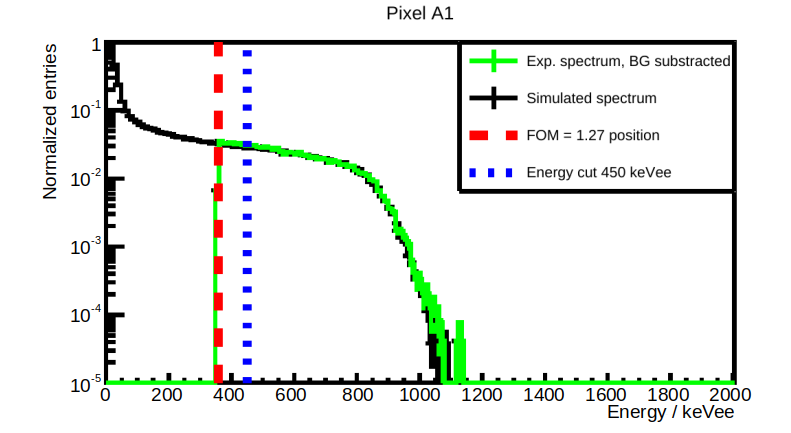}
    ~
    \includegraphics[width=.49\textwidth]{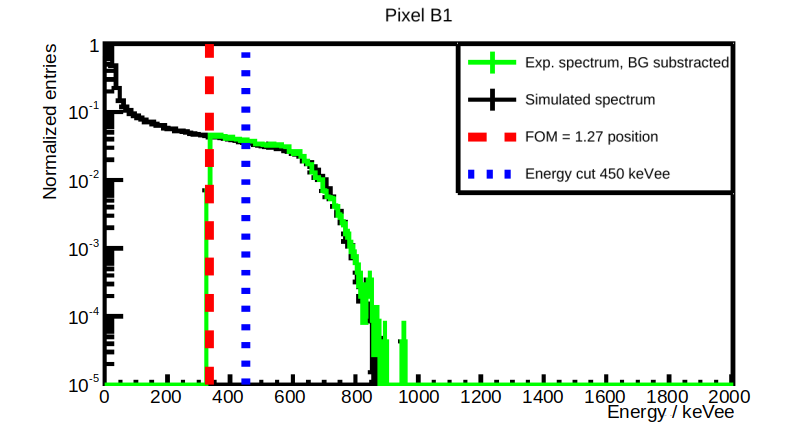}
    \caption{Example simulated (black) and measured (green) spectra, representing the two different stilbene classes. The red and blue dashed lines mark the experimental FOM=1.27 position and the artificial energy cut at \SI{450}{keVee}.  Left: Pixel A1. Right: Pixel B1. The energies are significantly smaller than for pixel A1.}
    \label{fig:sim_spec}
\end{figure}

One aspect of the simulation is the amount of scintillation light, which can be transferred to an energy spectrum in keVee by dividing the number of photons produced by stilbene's light yield. When excluding scintillation light caused by electrons and gammas, one gets a spectrum that can be compared to the measured one. In Figure \ref{fig:sim_spec}, very good agreement between simulation and experiment can be seen for two different pixels. In order to achieve this agreement, the simulation parameters \textit{Birk's constant, kB,} and the \textit{intrinsic resolution} were optimized. A larger value of $kB$ means that proton recoils lead to less light compared to electron interactions. The intrinsic resolution describes the fluctuation of the amount of created scintillation light. To get a good agreement for all pixels, it turned out that at least two different classes of stilbene need to be defined, one with a Birk's constant of $kB=0.135~\si{mm/MeV}$ and one with $kB=0.185~\si{mm/MeV}$ (red or blue stilbene crystals in Figure \ref{fig:geant4model}). These values of $kB$ cannot be seen as a physical Birk's constant but are simply parameters used to increase the simulation accuracy. The same is true for the intrinsic resolution. Reasons for the pixel to pixel differences may include the anisotropy of stilbene \cite{anisotropy}, differences in the light propagation within the scintillators, and variation of light detection at the SiPMs. Light propagation is not included explicitly in the simulation, so its effect is instead represented by the tuning of the aforementioned parameters. 

\noindent To calculate the simulated detection efficiency, the experimentally determined $FOM=1.27$ values are applied to the spectra on a pixel by pixel basis and the integral of the remaining spectra is calculated. Also, a simulation with a PE block was performed and its resulting spectrum subtracted. By dividing the integrals by the expected incoming amount of neutrons at the detector, calculated from the amount of created neutrons, their angular distribution and the distance to the detector, the efficiency is obtained. For the optimized simulation setup, a mean detection efficiency of $(9.87 \pm 0.01)\%$ in the forward direction was determined. This is approximately $0.3$ percentage points higher than the experimentally determined value but within the overall experimental uncertainty. When applying a \SI{450}{keVee} threshold instead of the $FOM=1.27$ thresholds to the simulated data, the mean efficiency is at $(6.25\pm0.01)\%$ approximately 0.9 percentage points larger than the experimental value.\\
In the \SI{90}{\degree} direction, however, the simulated efficiency is only $(9.36 \pm 0.01)\% $, which is approximately 2.5 percentage points smaller than the experimental one, and particularly smaller than in the forward direction. One reason for the higher experimental efficiency is the behavior of pixels A2 and A3, where the experimental efficiencies have quite large uncertainties compared to the other pixels, while the simulation uncertainties are similar for all pixels. This influences the weighted mean efficiency. However, also for the other pixels, the simulated efficiencies are around 1.5-2 percentage points smaller. When applying the \SI{450}{keVee} threshold instead, the mean efficiency is $(4.70\pm0.01)\%$, which is very close to the experimental result. It is therefore still an open question where the discrepancy for the FOM criterion comes from. Also, to get a good agreement between the simulated and measured spectra, the values of the $kB$ parameters need to be set to \SI{0.125}{mm/MeV} and \SI{0.175}{mm/MeV}, indicating a tendency to larger pulse heights at the same amount of deposited energy than for the forward direction measurement. It is an open question what causes these differences. It could be a temperature effect in the measurements, e.g. a higher SiPM overvoltage, even though a temperature correction is applied to the SiPM bias voltage. The SiPM temperature during the \SI{90}{\degree} measurements was approximately \SI{2}{\celsius} lower than during the forward direction measurements. 

\subsubsection{Simulated macroscopic cross sections}
\begin{figure}[htbp!]
    \centering
    \includegraphics[width=.9\textwidth]{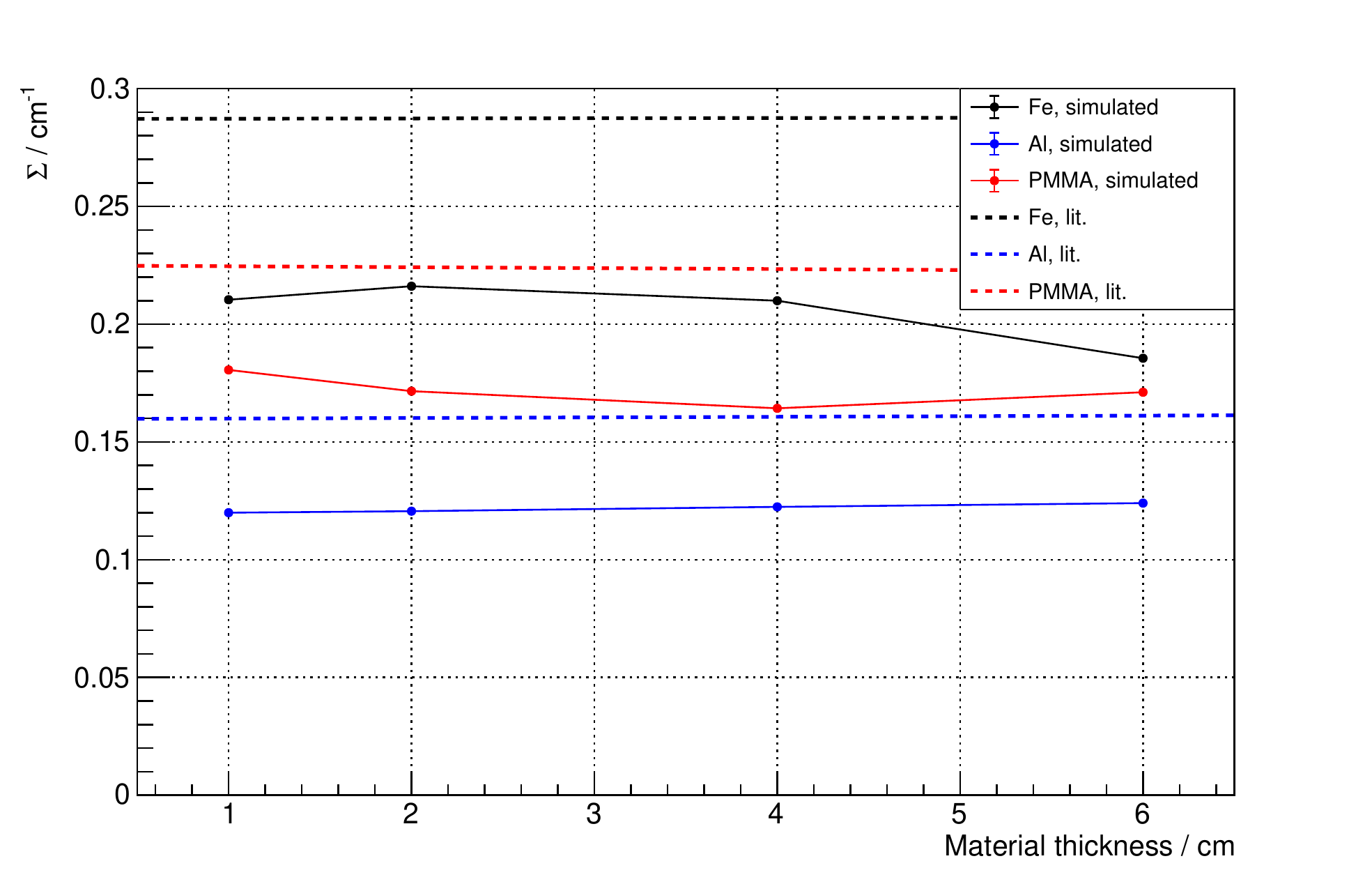}
    \caption{Simulated macroscopic cross section determined from the Fe-Al-PMMA step wedge's neutron attenuation. The literature cross sections are added for comparison as dotted lines.}
    \label{fig:sim_mac}
\end{figure}

Figure \ref{fig:sim_mac} shows the simulated macroscopic cross section $\Sigma$ determined from a simulation with the Fe-Al-PMMA step wedge which assumes a perfect alignment of the step wedge in front of the pixels and the detectors with respect to the source. $\Sigma$ is determined as in the experiment. \\
As in the experiment, the calculated cross sections are significantly smaller than the literature values for all three materials. Two possible reasons were identified in the simulation. The first contribution is from neutrons interacting in both the object and the detector, which cause up to approximately 8\% of the detected neutrons in the simulation. However, this effect alone cannot fully describe the discrepancy between simulated and literature cross sections. Another contribution comes from multiple interactions in the detector, either in two or more different pixels or in the detector housing and then in a pixel, leading to a background rate of neutrons. The background generated in this way reduces the determined cross section and is of relatively greater importance for pixels covered with a higher-attenuating material and/or a greater material thickness. This effect is not corrected by subtracting the PE measurement, as the PE block is much larger than the detector volume. Multiple interaction events in detector pixels with at least one interaction above the required FOM threshold occur in about 20\% - 25\% of the simulated events when assuming the mean FOM threshold of \SI{318}{keVee} in the forward direction. However, this value also includes very small energy losses of a few keV and does not consider if the neutron was scattered into the pixel where it was counted or vice versa. Using e.g. an arbitrarily chosen \SI{100}{keV} threshold, which corresponds to approximately 3.5\% of the initial neutron energy, the multi-pixel interaction fraction is already reduced to approximately 13\%.  Additionally, neutrons that interact in the detector box and then in the detector need to be considered. It would be interesting for the future to extend the simulation so that the order of neutron interactions in the detector can be reconstructed to quantify the influence of multiple interactions on the determined macroscopic cross section.

\begin{figure}[htbp!]
    \centering
    \begin{subfigure}{.49\textwidth}
    \includegraphics[width=\textwidth]{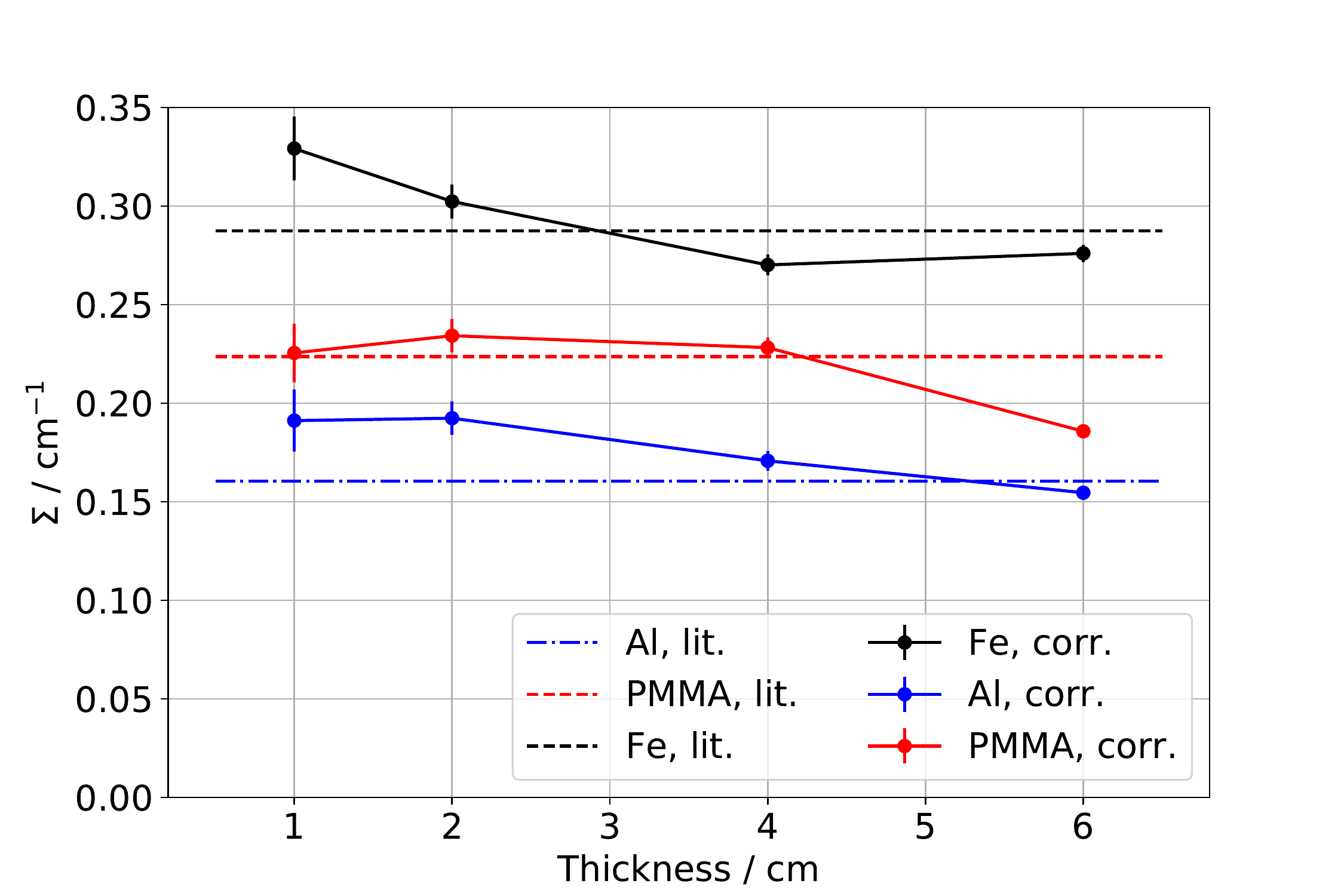}
    \subcaption{No shift.}
    \end{subfigure}
    ~
    \begin{subfigure}{.49\textwidth}
    \includegraphics[width=\textwidth]{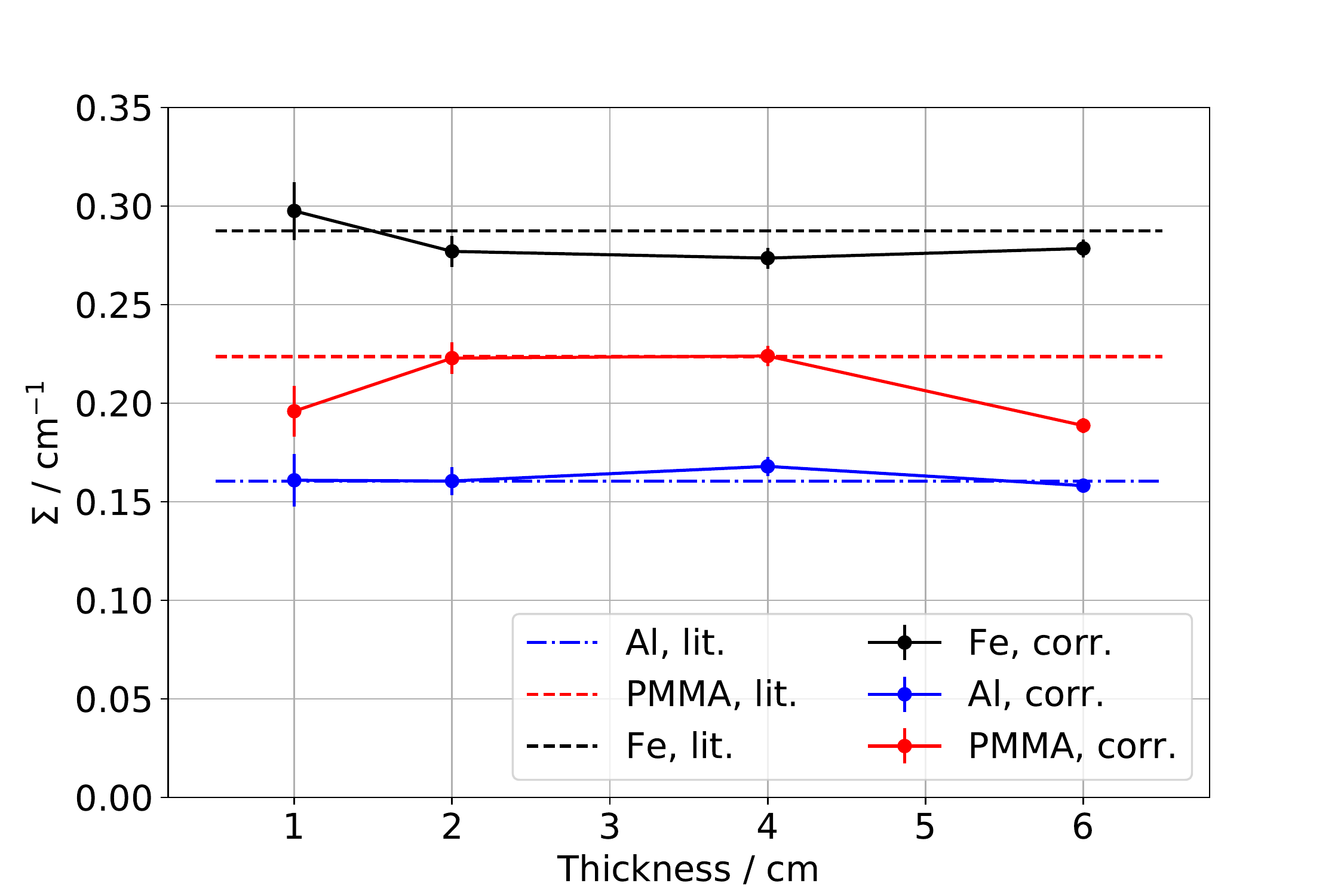}
    \subcaption{Vertical shift by \SI{1}{mm}.}
    \end{subfigure}
    \caption{Corrected experimental macroscopic cross sections by the simulated cross sections. (a): Using data from the unshifted simulation. (b): Using data from the vertically shifted simulation by \SI{1}{mm} upwards. }
    \label{fig:corrected_xs}
\end{figure}

\noindent Both the cross sections for small material thicknesses and the slope of the decrease of the cross sections towards higher material thicknesses are smaller in the simulation than in the experiment. A possible explanation could be a slight misalignment of the step wedge in the experiment. This could either be a shift or a rotation of the step wedge compared to the direct neutron direction of flight, influencing the attenuation in front of the detector pixels. To investigate effects of shifts, the step wedge in the simulation was misaligned in the vertical and horizontal direction by $\pm 1~\si{mm}$ and $\pm~2~\si{mm}$. It should be mentioned that a \SI{2}{mm} shift in the experiment is quite unlikely as great effort was made to correctly position the step wedge in front of the detector. \\
A misalignment in the horizontal direction to the right, shifting the PMMA part slightly into the empty detector row, can be limited to (much) less than \SI{1}{\milli\meter} as this misalignment would cause a much larger shadowing of up to approximately $-12\%$ in the empty row than observed in the measurement. Also, a shift of \SI{2}{mm} to the left can be excluded, as this would cause a smaller experimental cross section for PMMA than for Al, which is not observed.
The decrease of the cross section towards higher material thicknesses could be explained, among other possibilities, by a shift upwards. For example, a pixel that was assumed to be covered by \SI{1}{cm} of material would now be partially covered by \SI{2}{cm} of the same material. Such a shift leads to a higher experimental cross section. This effect is in percentage terms larger for the small material thicknesses. A simulation with such a shift of \SI{1}{mm} fits better to the data than the unshifted simulation. To illustrate this, correction factors for the cross section can be calculated from the simulated data by dividing the literature macroscopic cross sections by the simulated ones. When multiplying the experimental values by the correction factor, the resulting values should ideally correspond to the literature cross sections. Figure \ref{fig:corrected_xs} shows the results for the original and the vertically shifted case. The shifted simulation leads to better agreement between corrected values and literature values, even though there are still some deviations (e.g. \SI{6}{cm} of PMMA). However, it can be concluded that lower experimental macroscopic cross sections are expected, and the exact dependency of the thickness is probably explained by alignment difficulties.

%% file: conclusion_revised.tex
The 16 pixel detector prototype presented in this paper was successfully tested with a D-D neutron generator with mean emission energies between 2.3 and 2.8~MeV, approximately.  The detector's scintillation material stilbene allows neutron-gamma separation using pulse shape discrimination techniques. Applying these techniques, a neutron detection efficiency, averaged over all pixels, between $(9.11\pm 0.11) \%$ and $(11.95 \pm 0.12) \%$ was determined, applying a conservative lower energy cut by requiring $FOM>1.27$. A higher detection efficiency would be achievable by requiring a less conservative $FOM$ criterion with the disadvantage of a (slightly) larger contamination e.g. of neutron counts by gamma events and difficulties in the comparison with simulations, as neutron events need to be filtered if they lie in a region where they are mixed up with gamma events. This would be interesting if the measurement time needs to be reduced or a weaker neutron source shall be used. 
The detection efficiency was determined for three different angles relative to the generator's ion beam direction, i.e. three different mean neutron energies. Here, it was seen that the efficiency in the \SI{90}{\degree} direction was larger than in the \SI{0}{\degree} direction,  which is an unexpected result. Further studies should be done to explain this discrepancy.\\ 
With radiographic attenuation measurements of three different objects, it was shown that the detector setup allows distinguishing different materials and also the presence of material thickness differences of only \SI{0.5}{cm} for measurement times of approximately \SI{0.5}{h} to \SI{1.0}{h} by calculating a differential rate between measurements with and without object. The accuracy could possibly be improved by increasing the measurement time.  However, if material and thickness were unknown, in this radiographic measurement approach it would be impossible to accurately identify simultaneously material composition and thickness as there would be too many free variables. \\
From the differential rates, the macroscopic cross section for the attenuators' materials was calculated, showing a systematic underestimation compared to literature values. By using detailed Geant4 simulations, two main contributions for these deviations were identified: neutrons that interact in the attenuator but still arrive at the detector and multiple interactions in different detector pixels or in the detector housing and then in a pixel. Additionally, a possible misalignment of the step wedge was investigated by comparing the simulated and the measured macroscopic cross section values.  It would be interesting to further study the effect of multiple interactions on the determined cross sections using the simulations.